\documentclass[vecphys]{svmult}

\usepackage{makeidx}         
\usepackage{graphicx}        
\usepackage[bottom]{footmisc}

\def\lamb#1#2{$^{#1}_{\varLambda}${#2}}
\def\lam#1#2{$^{#1}_{~\varLambda}${#2}}
\def\Kpi{$({\mathrm K}^-\!,\upi^-\!)$ }
\def\Kpig{$({\mathrm K}^-\!,\upi^- \ugamma)$ }
\def\piK{$(\upi^+\!,{\mathrm K}^+\!)$ }
\def\piKg{$(\upi^+\!,{\mathrm K}^+\ugamma)$ }
\def\vlam{\varLambda}
\def\vsig{\varSigma}
\def\vdel{\varDelta}

\makeindex             

\begin{document}

\title*{Hypernuclear Gamma-Ray Spectroscopy and the Structure of
p-shell Nuclei and Hypernuclei}

\titlerunning{Hypernuclear Gamma-Ray Spectroscopy}

\author{D.J. Millener}

\institute{Brookhaven National Laboratory, Upton, NY 11973, USA}

%
%
\maketitle

\begin{abstract}
Information on \lamb{7}{Li}, \lamb{9}{Be}, \lam{10}B, \lam{11}{B},
\lam{12}{C}, \lam{15}{N}, and \lam{16}{O} from the Ge detector
array Hyperball is interpreted in terms of shell-model calculations
that include both $\vlam$ and $\vsig$ configurations with p-shell
cores. It is shown that the data puts strong constraints on the
spin dependence of the $\vlam$N effective interaction.
\footnote{Springer Lecture Notes in Physics 724, 31-79 (2007)}
\end{abstract}

\setcounter{minitocdepth}{2}
\section*{Contents}
\dominitoc

\section{Introduction}
\label{sec:introduction}

 The structure of $\vlam$ hypernuclei -- i.e. many-body systems
consisting of neutrons, protons, and $\vlam$ particles --
is an interesting subject in its own right. However, the finer details
of the structure of single-$\vlam$ hypernuclei, particularly the energy
spacings of doublets formed by the coupling of a $\vlam$ in the
lowest s orbit to a nuclear core state with non-zero spin, provide
information on the spin dependence of the effective $\vlam$N
interaction. This is important because data on the free YN interaction
are very sparse and essentially limited to spin-averaged s-wave scattering.

  The spectroscopy of $\vlam$ hypernuclei has been reviewed recently
by Hashimoto and Tamura~\cite{hashtam06}. The `workhorse' reactions used
to produce $\vlam$ hypernuclei have been the \Kpi (strangeness exchange)
and \piK (associated production) reactions that convert a neutron into a
$\vlam$. The elementary ${\rm n}({\rm K}^-\!,\upi^-\!)\vlam$ and
${\rm n}(\upi^+\!,{\rm K}^+\!)\vlam$
reactions have predominantly non-spin-flip character at the
incident beam energies used.

 The first information on $\vlam$ hypernuclei came from their production
via K$^-$ mesons stopped in emulsion followed by their $\upi^-$-mesonic
weak decay~\cite{davis86}. These studies provided  $\vlam$ separation
energies (B$_\vlam$ values) up to $\mathrm{A}\sim 15$. These could be
accounted for by a $\vlam$-nucleus potential of Woods-Saxon shape with
a depth of about 30\,MeV. A number of ground-state spins were determined
from angular correlation studies and weak-decay branching ratios, $\ugamma$
rays from excited states
of \lamb{4}{H} and  \lamb{4}{He} were seen, and so was proton
emission from excited states of \lam{12}{C}. Currently, counter
experiments with stopped K$^-$ mesons are being performed at
Frascati~\cite{agnello05}.

 The momentum transfer to the hypernucleus is rather small in the forward
direction in \Kpi reactions near the beam momenta of $\sim 800$\,MeV/c
used at CERN and BNL and the cross
sections for $\Delta L =0$ transitions are large. Because the cross
sections are proportional to the spectroscopic factors for neutron removal
from the target, the cross sections are largest when a high-spin
neutron orbit is just filled and the produced $\vlam$ occupies the same
orbit. Such transitions have been observed in selected nuclei up
to $^{209}$Bi (see~\cite{hashtam06}).

 The \piK reaction has been used at $p_\pi \!=\! 1.05$\,GeV/c where the
elementary cross section peaks strongly. The momentum transfer is
high ($q\sim 350$\,MeV/c) and the reaction selectively populates high-spin
states. The cross sections are smaller than for the \Kpi reaction
but the count rates for producing $\vlam$ hypernuclei can be comparable
because more intense pion beams can be used. Transitions from nodeless
high-spin neutron orbits can populate the full range of nodeless
bound, and just unbound, $\vlam$ orbitals and have been used to map
out the spectrum of $\vlam$ single-particle energies for selected nuclei
up to $^{208}$Pb (see~\cite{hashtam06}). This information has provided
a rather precise characterization of the $\vlam$-nucleus potential.
The best resolution, obtained at KEK using the SKS spectrometer
and a thin $^{12}$C target, is 1.45\,MeV~\cite{hotchi01}.

 The free $\vlam$ decays into a nucleon and a pion with a lifetime of
263\,ps. In a hypernucleus, the low-energy nucleon produced via this
decay mode is Pauli blocked and the process $\vlam$N$\to$NN rapidly
dominates with increasing mass number. Nevertheless, the measured weak
decay lifetimes of $\vlam$ hypernuclei remain around 200\,ps. This
means that particle-bound excited states of $\vlam$ hypernuclei
normally decay electromagnetically. Then it is possible to make use
of the excellent resolution of $\ugamma$-ray detectors to measure the
spacings of hypernuclear levels. The earliest measurements were made
with NaI detectors but the superior (few keV) resolution of Ge
detectors has been exploited since 1998 in the form of the Hyperball,
a large-acceptance Ge detector array. A series of experiments
on p-shell targets has been carried out at KEK and BNL using the \piKg
and \Kpig reactions, respectively~\cite{hashtam06}. As well as
$\ugamma$-ray transitions between bound states of the primary
hypernucleus, $\ugamma$-ray transitions are often seen from
daughter hypernuclei formed by particle emission (most often a proton)
from unbound states of the primary hypernucleus.

 The results of these $\ugamma$-ray experiments are interpreted in
terms of nuclear structure calculations with a parametrized effective YN
interaction as input.

\section{The $\vlam$N (YN) Effective Interaction}
\label{sec:YN}

\begin{figure}[t]
\centering
\includegraphics[width=11.0cm]{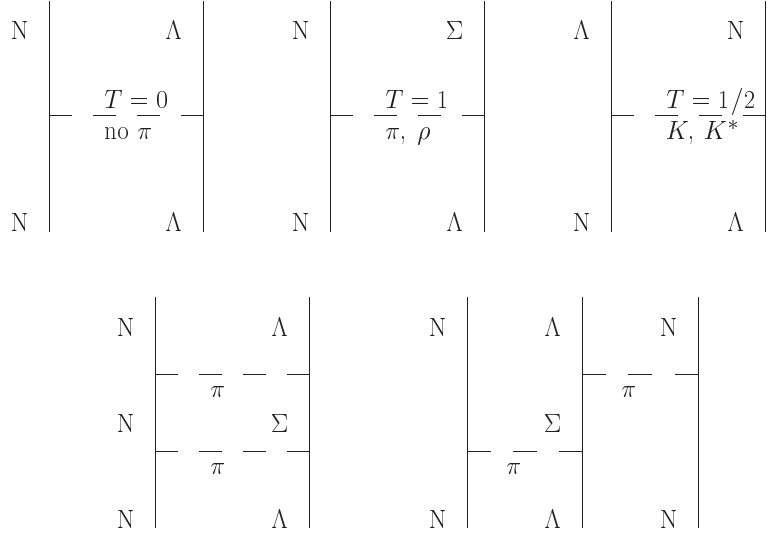}
\caption{Diagrams showing the important features
of the coupled $\varLambda\mathrm{N} - \varSigma\mathrm{N}$ strangeness $-1$
interaction for isospin 1/2. The last diagram shows the
two-pion exchange three-body interaction}
\label{fig:1}
\end{figure}

 The hyperon-nucleon interaction involves the coupled $\vlam$N and
$\vsig$N channels, as illustrated in Fig.~\ref{fig:1}. The
diagrams in Fig.~\ref{fig:1} make the point that the direct
$\vlam$N--$\vlam$N interaction does not contain a one-pion exchange
contribution because of isospin conservation (except for
electromagnetic violations via $\vlam$--$\vsig^0$ mixing) while the
coupling between the $\vlam$N and $\vsig$N channels does. For this
reason, the $\vlam$N interaction is relatively weak and there is
reason to believe that the three-body interaction in a hypernucleus
could be relatively important.

The free-space interactions are obtained as extensions of
meson-exchange models for the NN interaction by invoking, e.g., a
broken flavor SU3 symmetry. The most widely used model is the
Nijmegen soft-core, one-boson-exchange potential model known as
NSC97~\cite{rijken99}. The six versions of this model, labelled
NSC97a..f, cover a wide range of possibilities for the strength of
the central spin-spin interaction ranging from a triplet
interaction that is stronger than the singlet interaction to the
opposite situation. An extended soft-core version (ESC04) has
recently been published~\cite{rijken06}. Effective interactions
for use in a nuclear medium are then derived through a G-matrix
procedure~\cite{rijken99,rijken06}.

 The $\vlam$N effective interaction can be written (neglecting a
quadratic spin-orbit component) in the form
\begin{equation}
V_{\varLambda {\rm N}}(r) = V_0(r) +V_{\sigma}(r)~ \vec{s}_{\rm N}\cdot
 \vec{s}_{\varLambda} +  V_{\varLambda }(r)~\vec{l}_{\rm N\varLambda }\cdot
\vec{s}_{\varLambda}  + V_{\rm N}(r)~\vec{l}_{\rm N \varLambda }\cdot
\vec{s}_{\rm N}\nonumber +  V_{\rm T}(r)~S_{12}\; ,
\label{eq:vlam}
\end{equation}
where $V_0$ is the spin-averaged central interaction, $V_\sigma$ is the
difference between the triplet and singlet central interactions, $V_\vlam$
and $V_{\rm N}$ are the sum and difference of the strengths of the symmetric
spin-orbit (LS) interaction
$\vec{l}_{\rm N\Lambda }\cdot(\vec{s}_{\varLambda} +\vec{s}_{\rm N}$)
and antisymmetric spin-orbit (ALS) interaction
$\vec{l}_{\rm N\Lambda }\cdot(\vec{s}_{\varLambda} -\vec{s}_{\rm N}$),
and $V_{\rm T}$ is the tensor interaction with ($C^2$ is a normalized spherical
harmonic)
\begin{eqnarray}
S_{12} & = & 3(\vec{\sigma}_{\rm N}\cdot\hat{\vec{r}})(\vec{\sigma}_{\varLambda}
\cdot\hat{\vec{r}})-\vec{\sigma}_{\rm N}\cdot\vec{\sigma}_{\varLambda} \nonumber \\
 & = & \sqrt{6}\,C^2(\hat{\vec{r}})\cdot[\vec{\sigma}_{\rm N},
\vec{\sigma}_\varLambda]^2 \; .
\label{eq:tensor}
\end{eqnarray}
For the $\vlam$ in an s orbit, $\vec{l}_{\rm N\Lambda }$ is proportional to
$\vec{l}_{\rm N}$~\cite{gsd}. The effective $\vlam$N--$\vsig$N and
$\vsig$N--$\vsig$N interactions can be written in the same way.

Each term of the potential in (\ref{eq:vlam}) can be written in the form
\begin{equation}
 V_k(r)\ {\cal L}^k\cdot {\cal S}^k =  V_k(r)\
(-)^k\widehat{k}[{\cal L}^k,\,{\cal S}^k]^0 \; ,
\label{eq:lsk}
\end{equation}
where $k$ is the spherical tensor rank of the orbital and spin operators
and $\widehat{k}^2 = 2k+1$.

 So-called YNG interactions, in which each term of the effective interaction
is represented by an expansion in terms of a limited number of Gaussians with
different ranges,
\begin{equation}
 V(r) = \sum_i v_i\, \E^{-r^2/\beta_i^2}\; ,
\label{eq:gauss}
\end{equation}
are often used for the central and spin-orbit components with the
following form used for the tensor component,
\begin{equation}
V_T(r) = \sum_i v_i\, r^2\, \E^{-r^2/\beta_i^2} \; .
\label{eq:r2gauss}
\end{equation}
For example, the G-matrix elements  from a nuclear matter
calculation have been parametrized in this way, in which case the
YNG interactions have a density dependence (through $k_{\rm F}$).

 Given the interaction in YNG, or some other, radial form, two-body matrix
elements that define the interaction for a shell-model calculation
can be calculated using a chosen set of single-particle radial
wave functions. In the following, the procedure is sketched for harmonic
oscillator radial wave functions in the case of equal mass particles.
There are techniques to calculate the two-body matrix elements
for any (e.g., Woods-Saxon) radial wave functions but the harmonic
oscillator case illustrates where the important contributions come
from and suggests ways in which the interaction can be parametrized
in terms of the radial matrix elements themselves.

 Separating the space and spin variables in (\ref{eq:vlam}) and (\ref{eq:lsk})
using (\ref{eq:cprodsu2})
\begin{eqnarray}
 \langle l_1l_2LS|V|l_1'l_2'L'S'\rangle^{JT} & = & \sum_k (-)^{L'+S+J}
\left\{ \begin{array}{ccc} L & L' & k \\ S' & S & J \end{array}\right\}
\nonumber\\
& ~\times & \widehat{L}\langle l_1l_2L||V_k(r){\cal L}^k||l_1'l_2'L'\rangle
\widehat{S}\langle S||{\cal S}^k||S'\rangle \; .
\label{eq:vsep}
\end{eqnarray}

\begin{table}[b]
\centering
\caption{Two-particle reduced matrix elements of orbital and spin operators}
\label{tab:rme}
\begin{tabular}{lccccc}
\hline\noalign{\smallskip}
   & 1 & $\vec{s}_{\rm N}\cdot\vec{s}_\varLambda$ & LS & ALS & Tensor \\
\noalign{\smallskip}\hline\noalign{\smallskip}
$\widehat{S}\langle S||{\cal S}^k||S'\rangle$ & ~$\widehat{S}$ & ~$\widehat{S}
[2S(S+1)-3]/4$ &
$\delta_{SS'}\widehat{S}\sqrt{2}$ & ~$(-)^S(1-\delta_{SS'})\sqrt{3}$ &
$\widehat{S}\sqrt{20/3}$ \\
$\langle l||{\cal L}^k||l'\rangle$ & 1 & 1 & $\delta_{ll'}\sqrt{l(l+1)}$~ &
$\delta_{ll'}\sqrt{l(l+1)}$
& $\sqrt{6}\langle l020|l'0 \rangle$ \\
\noalign{\smallskip}\hline
\end{tabular}
\end{table}

 Harmonic oscillator wave functions have the unique property that a
transformation exists from the individual particle coordinates
$\vec{r}_1,\; \vec{r}_2$ to the relative and center of mass coordinates
$(\vec{r}_1-\vec{r}_2)/\sqrt{2},\; (\vec{r}_1+\vec{r}_2)/\sqrt{2}$ of
the pair~\cite{brody67}. This transformation and another application of
(\ref{eq:cprodsu2}) result in an expression in terms of radial integrals
in the relative coordinate $r=|\vec{r_1}-\vec{r_2}|$
\begin{eqnarray}
 \langle l_1l_2L||V_k(r){\cal L}^k||l_1'l_2'L'\rangle & = &
\sum_{N_{\rm c}L_{\rm c}ll'}
(-)^{l+L_{\rm c}+k+L'}\widehat{L'}\widehat{l}\left\{
\begin{array}{ccc} L_{\rm c} & l' & L' \\ k & L & l \end{array}\right\}
\nonumber\\
&~\times&\langle nlN_{\rm c}L_{\rm c},L|n_1l_1n_2l_2,L\rangle\langle
n'l'N_{\rm c}L_{\rm c},L'|
n_1'l_1'n_2'l_2',L'\rangle \nonumber \\
&~\times&\langle nl|V_k(r)|n'l'\rangle \langle l||{\cal L}^k||l'\rangle \; ,
\label{eq:vorb}
\end{eqnarray}
where the number of quanta associated with coordinate $r$ is given
by $q=2n+l$ ($n=0,1,..$) and energy conservation
$q_1+q_2 =q+Q_{\rm c}$ fixes $n$ (similarly $n'$).
The reduced matrix elements (see Appendix~\ref{sec:racsu2}) of the orbital
and spin operators are listed in Table~\ref{tab:rme}.
The radial integral can in turn be expressed in terms of Talmi integrals $I_p$
\begin{equation}
\langle nl|V(r)|n'l'\rangle = \sum_p\ B(nl,n'l';p)\, I_p \; .
\label{eq:vbnl}
\end{equation}
The harmonic oscillator radial relative wave function is a polynomial
in $r'$ times exp$(-r'^2)$ where $r'=|\vec{r_1}-\vec{r_2}|/\sqrt{2}\,b$
and $b^2 = \hbar/m\,\omega$. Then,
\begin{equation}
 I_p = \frac{2}{\Gamma(p+3/2)} \int_0^\infty r^{2p}\E^{-r^2}
V(\sqrt{2}rb)\,r^2\,\D r \; .
\label{eq:vip}
\end{equation}
For a Gaussian potential, $V(r) = V_0\, \mathrm{exp}(-r^2/\mu^2)$,
with $\theta = b/\mu$,
\begin{equation}
 I_p = \frac{V_0}{(1+2\,\theta^2)^{p+3/2}} \; .
\label{eq:vipgauss}
\end{equation}

 For the case of a $\vlam$ in an s orbit attached to a light nucleus,
the expressions for the matrix elements of each component of the
interaction are shown in Table~\ref{tab:vlam}. In this simple case, the
$I_p$ are equal to the relative matrix elements in the angular momentum
states denoted by $p$ (the superfluous superscripts denote the
interaction in even or odd relative states). For the nuclear p shell,
there are just five $p_{\rm N}s_\vlam$ two-body matrix elements formed from
$p_{1/2}s_{1/2}(0^-,1^-)$ and $p_{3/2}s_{1/2}(1^-,2^-)$ (alternatively,
$\mathrm{L}=1$ and $\mathrm{S}=0,1$). This means that the five radial
integrals $\overline\mathrm{V}$, $\vdel$, S$_\vlam$, S$_\mathrm{N}$, and T
associated with each operator in  Table~\ref{tab:vlam}
can be used to parametrize the $\vlam$N effective interaction. In
Appendix~\ref{sec:twobod}, the $p_{\rm N}s_\varLambda$ two-body matrix elements
are given in terms of the parameters in both LS and $jj$ coupling.

\begin{table}
\centering
\caption{$\vlam$N (YN) interaction parameters}
\label{tab:vlam}
\begin{tabular}{cccc}
\hline\noalign{\smallskip}
 $V_{\rm N\Lambda}$ &  ~~$s_{\mathrm N}s_\varLambda$~~ &  $p_{\rm N}s_\varLambda$ &
\lamb{7}{Li} values (MeV) \\
\noalign{\smallskip}\hline\noalign{\smallskip}
$V_0$ &  $I^e_0$ & $\overline\mathrm{V} = \frac{1}{2}(I^e_0 + I^o_1)$ &
($-1.22$) \vspace{1.5pt}\\
$V_\sigma\vec{s}_{\rm N}\cdot\vec{s}_\varLambda$ &  $I^e_0$  & $\vdel = \frac{1}{2}
(I^e_0 + I^o_1)$ &   \phantom{$-$}0.480  \vspace{1.5pt}  \\
  $V_\varLambda\,\vec{l}_{\rm N}\cdot\vec{s}_\varLambda$ &  &  S$_\vlam
= \frac{1}{2} I^o_1$ &  $-0.015$ \vspace{1.5pt}\\
 $V_{\rm N}\,\vec{l}_{\rm N}\cdot\vec{s}_{\rm N}$ &  & S$_\mathrm{N} =
\frac{1}{2} I^o_1$ &  $-0.400$ \vspace{1.5pt}\\
  $V_{\rm T}\,S_{12}$ &  & $\mathrm{T} = \frac{1}{3} I^o_1$  & \phantom{$-$}0.030 \\
\noalign{\smallskip}\hline
\end{tabular}
\end{table}

 A comprehensive program for the shell-model analysis of $\vlam$ binding
energies for p-shell hypernuclei was set out by Gal, Soper, and
Dalitz~~\cite{gsd}, who also included the three-body
double-one-pion-exchange $\vlam$NN interaction shown in Fig.~\ref{fig:1}.
This interaction does not depend on the spin of the $\vlam$ and
was characterized by a further five radial integrals. Dalitz and
Gal went on to consider the formation of p-shell hypernuclear
states via \Kpi and $(\,{\rm K}^-,\upi^0)$ reactions and the prospects for
$\ugamma$-ray spectroscopy based on the decay of these states~\cite{dg78}.
Unfortunately, knowledge of the ground-state B$_\vlam$ values plus a few
constraints from known ground-state spins was insufficient to provide
definitive information on the spin-dependence of the $\vlam$N interaction.

 The most direct information on the spin dependence of the $\vlam$N
effective interaction comes from the spacing of $s_\vlam$ doublets based
on core states with non-zero spin. These spacings depend on the parameters
$\vdel$, S$_\vlam$, and T that are associated with operators that involve
the $\vlam$ spin. The energy separations of states based on different
core states depend on S$_\mathrm{N}$, but these separations can also
be affected by the three-body interaction. Millener, Gal, Dover, and
Dalitz~\cite{mgdd85} made estimates for $\vdel$, S$_\vlam$, S$_\mathrm{N}$,
and T using new information, particularly on $\ugamma$-ray transitions
in \lamb{7}{Li} and \lamb{9}{Be}~\cite{may83}, together with theoretical
input from YN interaction models. These estimates were close to the
values given in the right-hand column of Table~\ref{tab:vlam} (the
bracketed value for $\overline\mathrm{V}$ is not fitted) which fit the
now-known energies of the four bound excited states of \lamb{7}{Li} (see
Sect.~\ref{sec:L7Li}). An alternative set of parameters was proposed
by Fetisov, Majling, \v{Z}ofka, and Eramzhyan~\cite{fmze91} who
were motivated by the non-observation of a $\ugamma$-ray transition
from the ground-state doublet of \lam{10}{B} in the first experiment
using Ge detectors~\cite{chrien90}.

 Experiments with the Hyperball, starting in 1998 with
$^7$Li\,\piKg\lamb{7}{Li} at KEK and $^9$Be\,\Kpig\lamb{9}{Be} at BNL,
have provided the energies of numerous $\ugamma$-ray transitions,
together with information on relative intensities and lifetimes.
The progress of the theoretical interpretation in terms of shell-model
calculations has been summarized at HYP2000~\cite{millener01} and
HYP2003~\cite{millener05}. By the latter meeting, $\vsig$ degrees of freedom
were being included explicitly through the inclusion of both $\vlam$
and $\vsig$ configurations in the shell-model basis.

 The most convincing evidence for the importance of $\vlam$--$\vsig$
coupling comes from the s-shell hypernuclei and this is described
in the following section. This is followed by a discussion of
\lamb{7}{Li} in Sect.~\ref{sec:L7Li}. Because the LS structure of
the p-shell core nuclei plays an important role in picking out
particular combinations of the spin-dependent $\vlam$N parameters,
Sect.~\ref{sec:pshell} is devoted to a general survey of p-shell
calculations, spectra, and wave functions. This information is used in
subsequent sections that are devoted to the remaining hypernuclei,
up to \lam{16}{O}, for which data, particularly from $\ugamma$ rays, exists.

\section{The s-shell $\vlam$ Hypernuclei}
\label{sec:s-shell}

 The data on the s-shell hypernuclei is shown in Table~\ref{tab:sshell}.
The B$_\vlam$ values come from emulsion data~\cite{davis86} and the
excitation energies from $\ugamma$ rays observed following the stopping
of negative kaons in $^6$Li and $^7$Li~\cite{bedjidian79}.

\begin{table}
\centering
\caption{Data on the s-shell $\vlam$ hypernuclei}
\label{tab:sshell}
\begin{tabular}{ccccc}
\hline\noalign{\smallskip}
Hypernucleus &  $J^\pi (gs)$ & B$_\vlam$ (MeV) & $J^\pi$ & E$_x$ (MeV) \\
\noalign{\smallskip}\hline\noalign{\smallskip}
 \lamb{3}{H} & $1/2^+$  &  0.13(5) & & \vspace{1.3pt}\\
 \lamb{4}{H} &  $0^+$  & 2.04(4) &   $1^+$ &  1.04(5)\vspace{1.3pt} \\
 \lamb{4}{He} &  $0^+$ & 2.39(3) &   $1^+$ &  1.15(4)\vspace{1.3pt} \\
 \lamb{5}{He} &  $1/2^+$   & 3.12(2) & & \\
\noalign{\smallskip}\hline
\end{tabular}
\end{table}

 The spin-spin component of the $\vlam$N interaction contributes to
the splitting of the $1^+$ and $0^+$ states of \lamb{4}{H} and \lamb{4}{He}.
In the case of simple $s^3s_\vlam$ configurations, the contribution
is given by the radial integral of the spin-spin interaction in s
states (the $\varDelta$ in Table~\ref{tab:vlam}) using
\begin{eqnarray}
\label{eq:spin}
\sum_i\vec{s}_i\cdot \vec{s}_\vlam & = & \vec{S}_{\rm c}\cdot \vec{s}_\vlam
\nonumber \\
 & = & \frac{1}{2}[\vec{S}^2 -\vec{S}_{\rm c}^2 - \vec{s}_\vlam^2]\; .
\end{eqnarray}
However, it has long been recognized as a problem to describe
simultaneously the binding energies of the s-shell hypernuclei with a
central $\vlam$N interaction and that this problem might be solved by
$\vlam$--$\vsig$ coupling which strongly affects the $0^+$ states
of the $\mathrm{A}\!=\!4$ hypernuclei. Recently, there has been a clear
demonstration of these effects and it was found that  the spin-spin
and $\vlam$--$\vsig$ coupling components of the NSC97e and NSC97f
interactions give comparable contributions to the $1^+$--$0^+$
doublet splitting~\cite{akaishi00}. Subsequent studies
using a variational method with Jacobi-coordinate Gaussian-basis
functions~\cite{hiyama01}, Faddeev-Yakubovsky calculations~\cite{nogga02},
and stochastic variational calculations with correlated
Gaussians~\cite{nemura02} have confirmed and illustrated various
aspects of the problem.

 Akaishi et al.~\cite{akaishi00} calculated G-matrices for a small model
space of s orbits only, writing two-component wave functions for
either the $0^+$ or the $1^+$ states of \lamb{4}{He} (or \lamb{4}{H})
with isospin $\mathrm{T}=1/2$
\begin{equation}
\label{eq:4he}
|{^4_\vlam\textrm{He}}\rangle = \alpha s^3s_\vlam + \beta s^3s_\vsig \; .
\end{equation}
The $\vsig$ component is 2/3 $\vsig^+$ and 1/3 $\vsig^0$ for \lamb{4}{He}
(2/3 $\vsig^-$ and 1/3 $\vsig^0$ for \lamb{4}{H}). The off-diagonal
matrix element
\begin{equation}
\label{eq:god}
  v(J) =\langle s^3s_\varLambda,J|V| s^3s_\varSigma,J\rangle
\end{equation}
can be derived from the $\vlam$N$-\vsig$N G matrix for $0s$ orbits,
where $V$ is used for the potential representing the G matrix interaction,
by splitting one nucleon off from the $s^3$ configurations using the
fractional parentage expansion
\begin{equation}
\label{eq:scfp}
 |s^3\rangle = \sum_{S(T)} \frac{1}{\sqrt{2}}(-)^{1+S}|[s^2(TS),s]
(1/2\,1/2)\rangle\; ,
\end{equation}
where $\mathrm{TS} = 0\,1$ or $1\,0$.  Coefficients
of fractional parentage (cfp) specify how to construct a fully
antisymmetric $n$-particle state from antisymmetric $(n-1)$-particle
states coupled to the nth particle [cf. (\ref{eq:cfp})]. In this simple
case, the magnitude of the cfp is determined
by the symmetry with respect to T and S and the phase
appears twice and cancels out in the problem at hand. Then, by recoupling
on either side of (\ref{eq:god}) (see Appendix~\ref{sec:racsu2}),
\begin{eqnarray}
v(J)&  = & \frac{3}{2}\;\sum_{S\bar{S}}U(S\,1/2\,J\,1/2,\,1/2\,\bar{S})^2
U(T\,1/2\,1/2\,0,\,1/2\,1/2)
\nonumber \\
 & & \times\ U(T\,1/2\,1/2\,1,\,1/2\,1/2)
\langle ss_\varLambda,\bar{S}|V|ss_\varSigma,\bar{S}\rangle \; ,
\label{eq:odme}
\end{eqnarray}
where the factor of 3 appears because any one of the three s-shell
nucleons can be chosen from an antisymmetric wave function.
Specializing to the case of $\mathrm{J}=0$
\begin{eqnarray}
 v(0) & = & \frac{3}{2}\,{^3V}-\frac{1}{2}\,{^1V} \nonumber \\
 & = & \overline\mathrm{V}+\frac{3}{4}\vdel \; ,
\label{eq:od0}
\end{eqnarray}
where
\begin{equation}
\overline\mathrm{V} =\frac{1}{4}\,{^1V}+\frac{3}{4}\,{^3V}\; ,
\quad\quad \vdel = {^3V}-{^1V}
\label{eq:vdel}
\end{equation}
Similarly,
\begin{eqnarray}
 v(1) & = & \frac{1}{2}\,{^3V}+\frac{1}{2}\,{^1V} \nonumber \\
 & = & \overline\mathrm{V}-\frac{1}{4}\vdel \; .
\label{eq:od1}
\end{eqnarray}

 Taking round numbers derived using the 20-range Gaussian
interaction of \cite{akaishi00} which represents NSC97f yields
$^3V\! = \!4.8$\,MeV
and $^1V\! = -1.0$\,MeV, which give $\overline\mathrm{V}\! = 3.35$\,MeV and
$\vdel\! = 5.8$\,MeV. Then, $v(0)\! = 7.7$\,MeV and $v(1)=1.9$\,MeV.
In a simple $2\times 2$ problem, the energy shifts of the
$\vlam$-hypernuclear states are given by $\sim v(J)^2/\Delta E$ with
$\Delta E\sim 80$\,MeV (and the admixture $\beta\sim -v(J)/\Delta E$).
Thus, the energy
shift for the $0+$ state is $\sim 0.74$\,MeV while the shift for the
$1^+$ state is small. The result is close to that for the NSC97f
interaction in Fig.~1 of \cite{akaishi00}.

 The same method can be used to obtain the singlet and triplet
contributions of the $\vlam$ interaction for all the s-shell hypernuclei
in the case of simple s-shell nuclear cores. The results are given in
Table~\ref{tab:sshellspin}. The expressions in terms of
$\overline\mathrm{V}$ and $\vdel$ can be written down by inspection.

\begin{table}
\centering
\caption{Contributions of singlet and triplet interactions to
s-shell hypernuclei}
\label{tab:sshellspin}
\begin{tabular}{cccc}
\hline\noalign{\smallskip}
Hypernucleus &  $J^\pi (gs)$ & $^3V$ and $^1V$ & $\overline\mathrm{V}$
and $\vdel$ \\
\noalign{\smallskip}\hline\noalign{\smallskip}
 \lamb{3}{H} & $1/2^+$  & ~~3/2\,$^1V+ 1/2\, ^3V$ &
$2\overline\mathrm{V}-\vdel$\vspace{1.1pt}\\
 \lamb{3}{H} & $3/2^+$  & 2\,$^3V$ & ~~$2\overline\mathrm{V}+1/2\vdel$
\vspace{1.1pt}\\
 \lamb{4}{He}, \lamb{4}{H} &  $0^+$  &  3/2\,$^1V+ 3/2\, ^3V$ &
$3\overline\mathrm{V} -3/4\vdel$\vspace{1.1pt}\\
 \lamb{4}{He}, \lamb{4}{H} &  $1^+$ & 1/2\,$^1V+ 5/2\, ^3V$ &
$3\overline\mathrm{V} +1/4\vdel$\vspace{1.1pt}\\
 \lamb{5}{He} &  $1/2^+$   &  $^1V+ 3\, ^3V$ & $4\overline\mathrm{V}$\\
\noalign{\smallskip}\hline
\end{tabular}
\end{table}

\section{The \lamb{7}{Li} Hypernucleus}
\label{sec:L7Li}

 The first p-shell hypernucleus with particle-stable excited states that
can be studied by $\ugamma$-ray spectroscopy is \lamb{7}{Li} and it is
of interest to compare the effects of the $\vlam$N spin-spin
interaction and $\vlam$--$\vsig$ coupling in \lamb{7}{Li} with those
in \lamb{4}{H} and \lamb{4}{He}.

 The low-lying states of \lamb{7}{Li}  consist of a $\vlam$
in an s orbit coupled (weakly) to a $^6$Li core. Only the $1^+;0$
($J^\pi;T$) ground
state of $^6$Li is stable with respect to deuteron emission but the
$\vlam$ brings in extra binding energy and the lowest particle-decay threshold
for \lamb{7}{Li} is \lamb{5}{He}$+{\rm d}$ at 3.94(4)\,MeV derived from
\begin{eqnarray}
\mathrm{S}_\mathrm{d}(^7_\varLambda\mathrm{Li}) & = & \mathrm{S}_\mathrm{d}
(^6\mathrm{Li}) +\mathrm{B}_\varLambda(^7_\varLambda\mathrm{Li})
-  \mathrm{B}_\varLambda(^5_\varLambda\mathrm{He})\nonumber \\
 & = & 1.475 + 5.58(3) - 3.12(2) \; ,
\label{eq:7Lid}
\end{eqnarray}
where the B$_\vlam$ values (errors in parentheses) come from emulsion
studies~\cite{davis86}.

\begin{figure}[b]
\centering
\includegraphics[width=11.8cm]{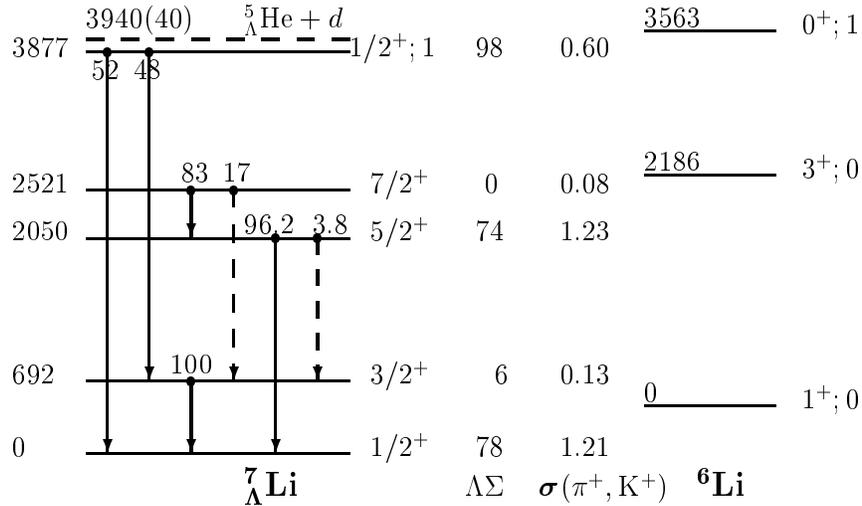}
\caption{The spectrum of \lamb{7}{Li} determined from experiments
KEK E419 and BNL E930 with the Hyperball
detector. All energies are in keV. The solid arrows denote observed
$\ugamma$-ray transitions. The $\ugamma$-ray branching ratios are theoretical
and the dashed arrows correspond to unobserved transitions.  For each state
of \lamb{7}{Li}, the calculated energy shifts due to $\vlam$--$\vsig$
coupling and the calculated relative populations via the \piK reaction
are given~\cite{hiyama99}. The core states of $^6$Li are shown on the right}
\label{fig:lli7}
\end{figure}

 Figure~\ref{fig:lli7} shows the spectrum of \lamb{7}{Li} determined from
experiments KEK E419~\cite{tamura00,tanida01} and BNL E930~\cite{ukai06} with
the Hyperball detector. The four $\ugamma$-rays seen in \cite{tamura00}
-- all except the $7/2^+\to 5/2^+$ transition -- show that the state
based on the $0^+;1$ state of $^6$Li is bound at an excitation energy of
3.88\,MeV. Only the $5/2^+\to 1/2^+$ transition was previously known from
an experiment at BNL with NaI detectors~\cite{may83}. Because the $3/2^+$
state is expected to be weakly populated in the \piK reaction, much of the
intensity of the 692-keV $\ugamma$-ray transition comes from feeding via
the $\ugamma$-ray transition from the $1/2^+;1$ level. The $7/2^+\to 5/2^+$
doublet transition was seen in $\ugamma\ugamma$ coincidence with the
$5/2^+\to 1/2^+$ transition following $^3$He emission from highly-excited
states of \lam{10}{B} (the s-hole region) produced via the \Kpi reaction on
$^{10}$B~\cite{ukai06}.

 Shell-model calculations for p-shell hypernuclei start with the
Hamiltonian (Y can be a $\vlam$ or a $\vsig$)
\begin{equation}
 H = H_{\rm N} + H_{\rm Y} + V_{\rm NY} \; ,
\label{eq:hamyn}
\end{equation}
where $H_{\rm N}$ is some empirical Hamiltonian for the p-shell core,
the single-particle $H_{\rm Y}$ supplies the $\sim 80$\,MeV mass difference
between $\vlam$ and $\vsig$, and $V_{\rm NY}$ is the YN interaction.
The two-body matrix elements of the YN interaction between states
of the form $(p_{\rm N}s_{\rm Y})$ can be parametrized in the form given in
Table~\ref{tab:vlam} (see Appendix~\ref{sec:twobod}). This form
applies to the direct $\vlam$N interaction, the $\vlam$N--$\vsig$N
coupling interaction, and the direct $\vsig$N interaction for both
isospin 1/2 and 3/2 (which is included in the calculations). The
shell-model basis states are chosen to be of the form
\begin{equation}
 |(p^n\alpha_{\rm c}J_{\rm c}T_{\rm c},j_{\rm Y}Y)JT\rangle\; ,
\label{eq:basis}
\end{equation}
where the hyperon is coupled in angular momentum and isospin
to eigenstates of the p-shell Hamiltonian for the core. This
is known as a weak-coupling basis and, indeed, the mixing of
basis states in the hypernuclear eigenstates is generally
very small. In this basis, the core energies are taken from
experiment where possible and from the p-shell calculation otherwise.

\begin{table}[b]
\centering \caption{Wave functions for $A\!=\!6$ using a number of
different interactions. The $2^+;0$ and $3^+;0$ states are
uniquely $^3$D} \label{tab:6liwfn}
\begin{tabular}{ccrrrrr}
\hline\noalign{\smallskip}
 $J_n^\pi$ & $^{(2S+1)}$L & fit69 & fit5 & CK616 & CK816 & CKPOT \\
\noalign{\smallskip}\hline\noalign{\smallskip}
 $1^+_1;0$ & & & & & & \\
 & $^3$S  & 0.9873 & 0.9906 & 0.9576 & 0.9484 & 0.9847 \\
 & $^3$D  & ~$-0.0422$ & ~$-0.0437$ & ~$-0.2777$ & ~$-0.3093$ & ~$-0.1600$ \\
 & $^1$P  & $-0.1532$ & $-0.1298$ & $-0.0761$ & $-0.0703$ & $-0.0685$ \\
 $1^+_2;0$  & & & & & &  \\
 & $^3$S &  0.0287 & $-0.0347$ & $-0.2810$ & $-0.3082$ & $-0.1426$ \\
 & $^3$D &  $-0.9007$ & $-0.9987$ & $-0.9591$ & $-0.9510$ & $-0.9673$ \\
 & $^1$P &  0.4334 & 0.0708 & $-0.0354$ & 0.0259 & 0.2096 \\
 $0^+_1;1$  & & & & & &  \\
 & $^1$S &  0.9560 & 0.9909 & 0.9997 & 0.9999 & 0.9946 \\
 & $^3$P &  0.2935 & 0.1348 & 0.0247 & $-0.0137$ & 0.1036 \\
 $2^+_1;1$  & & & & & & \\
 & $^1$D &  0.8760 & 0.9827 & 0.9486 & 0.9959 & 0.9839 \\
 & $^3$P &  0.4824 & 0.3148 & 0.1854 & 0.0905 & 0.1789 \\
\noalign{\smallskip}\hline
\end{tabular}
\end{table}

 For \lamb{7}{Li}, the basis states are of the form $|p^2s_\vlam\rangle$
and $|p^2s_\vsig\rangle$. The $p^2$ wave functions for $^6$Li are close
to the LS-coupling limit, as can be seen from Table~\ref{tab:6liwfn}
where wave functions are given for all three of Cohen and Kurath's
interactions~\cite{ck65} and two other interactions fitted to p-shell
data. As discussed in more detail in Sect.~\ref{sec:pshell}, the central
interaction is attractive in spatially even (S and D) states and repulsive
in (P) odd states. The $3^+;0$ state in Fig.~\ref{fig:lli7} is the lowest
member of an $\mathrm{L}\!=\!2$, $\mathrm{S}\!=\!1$ ($^3$D) triplet completed
by a $2^+;0$ state at 4.31\,MeV and a $1^+;0$ state at 5.65\,MeV; the
$2^+;1$ ($^1$D) state is at 5.67\,MeV. The $\mathrm{L}\!=\!1$ admixtures
are largely through the one-body spin-orbit interaction. The
$p_{1/2} - p_{3/2}$ splittings at $\mathrm{A}\!=\!5$
are small (0.14--1.29\,MeV) for the Cohen and Kurath interactions
(the p states are unbound at $\mathrm{A}\!=\!5$ and the $p_{1/2}$ energy is
poorly defined). The fit69 interaction has a larger splitting of
3.5\,MeV while the fit5 interaction, the one used in the hypernuclear
calculation, has an intermediate value of 1.8\,MeV.

 The structure of the core nucleus means that the $3/2^+$ and $7/2^+$
states are mainly $\mathrm{L}\!=\!0$, $\mathrm{S}\!=\!3/2$ and purely
$\mathrm{L}\!=\!2$, $\mathrm{S}\!=\!3/2$, respectively. This accounts for
their low population in the \piK reaction which is dominantly
non-spin-flip (the $^7$Li ground state has $\mathrm{L}\!=1$,
$\mathrm{S}\!=\!1/2$,  $\mathrm{J}\!=\!3/2$). The $1/2^+$ states are mainly
 $\mathrm{L}\!=\!0$, $\mathrm{S}\!=\!1/2$ while the $5/2^+$ state is 7/9
$\mathrm{S}\!=\!1/2$ and 2/9 $\mathrm{S}\!=\!3/2$ in the LS limit for the
core. In this limit, it is easy to derive the contribution of each
of the spin-dependent $\vlam$N parameters to the binding energies.

\begin{table}[b]
\centering
\caption{Contributions of the spin-dependent $\vlam$N terms to the
binding energies of the five bound states of \lamb{7}{Li} given
as the coefficients of each of the $\vlam$N effective
interaction parameters. In the $\vlam\vsig$ column the gains
in binding energy due to $\vlam$--$\vsig$ coupling are given
in keV (same as in Fig.~\ref{fig:lli7})}
\label{tab:L7Liabs}
\begin{tabular}{crrrrr}
\hline\noalign{\smallskip}
  $J^\pi_i;T$ & $\vlam\vsig$ & $\vdel$ & S$_\vlam$ & S$_\mathrm{N}$ & T \\
\noalign{\smallskip}\hline\noalign{\smallskip}
  $1/2^+;0$ & ~78 & ~$-0.975$  & ~$-0.025$ &  ~~0.242 &   0.080  \\
  $3/2^+;0$ &  6 &   0.486   &   0.013  &  0.253 & ~$-0.205$ \\
  $5/2^+;0$ & 74 & $-0.796$  & $-1.165$ &  0.980 &   1.177  \\
  $7/2^+;0$ &  0 &   0.500   &   1.000  &  1.000 & $-1.200$ \\
  $1/2^+;1$ & 98 & $-0.002$  &   0.002  &  0.453 & $-0.005$ \\
\noalign{\smallskip}\hline
\end{tabular}
\end{table}

 These contributions for the full shell-model calculation are given in
Table~\ref{tab:L7Liabs} as the coefficients of each of the $\vlam$N
effective interaction parameters. In the LS limit for the ground-state
doublet, only $\vdel$ contributes while for the excited-state doublet
all terms contribute. For the $1/2^+;1$ state, there would be no
contributions. However in the realistic case, S$_\mathrm{N}$ contributes
substantially for the predominantly $\mathrm{L}\! =\! 0$ cases. This is
because the associated operator $\vec{l}_\mathrm{N}\cdot\vec{s}_\mathrm{N}$
connects the $\mathrm{L}\!=\! 0$ and  $\mathrm{L}\! = \!1$ basis states of
the core giving rise to a linear dependence on the amplitude of the
$\mathrm{L}\! = \!1$ admixture. This
admixture is quite sensitive to the model for the p-shell core
(see Table~\ref{tab:6liwfn}). The hypernuclear shell-model states
are very close to the weak-coupling limit. For the $5/2^+$ state,
there is a 1.28\% admixture based on the the $2^+;0$ core state
(because the $2^+$ and $3^+$ core states share the same L and S).
Otherwise, the intensity of the dominant basis state is $>99.7$\%.

\begin{table}
\centering
\caption{Energy spacings in \lamb{7}{Li}. $\Delta E_\mathrm{C}$ is the
contribution of the core level spacing. The first line in each case
gives the coefficients of each of the $\vlam$N effective
interaction parameters as they enter into the spacing while the
second line gives the actual energy contributions to the spacing
in keV}
\label{tab:L7Li}
\begin{tabular}{crrrrrrr}
\hline\noalign{\smallskip}
  $J^\pi_i -J^\pi_f$ & $\Delta E_\mathrm{C}$ & $\vlam\vsig$ & $\vdel$
& S$_\vlam$ & S$_\mathrm{N}$ &  T & $\Delta E$   \\
\noalign{\smallskip}\hline\noalign{\smallskip}
  $3/2^+ - 1/2^+$ &  &   &  ~~1.461 & 0.038  &  ~~0.011 &  ~$-0.285$ & \\
  & 0 & 72   &  628  &   $-1$  &  $-4$ &   $-9$  &  693  \\
 $5/2^+ - 1/2^+$ &  &  & 0.179 & ~$-1.140$  &  0.738 &  1.097 & \\
 & ~~2186 & 4 &  77  & 17  & $-288$ & 33  &  2047  \\
$1/2^+ - 1/2^+$ &  &  & 0.972 & $-0.026$  &  0.211 &  $-0.085$ & \\
 & 3565 & ~$-20$ &  418  & 0  & $-82$ & $-3$  &  ~~3886 \\
$7/2^+ - 5/2^+$ &  &  &  1.294 & 2.166  &  0.020 &  $-2.380$ & \\
  & 0 & 74 &  557  & $-32$  & $-8$ & $-71$  &  494 \\
\noalign{\smallskip}\hline
\end{tabular}
\end{table}

 Table~\ref{tab:L7Li} re-expresses the same information in terms of
energy differences between states and gives the actual energy
contributions for the parameter set
\begin{equation}
\vdel= 0.430\quad \mathrm{S}_\vlam =-0.015\quad \mathrm{S}_\mathrm{N}
 = -0.390 \quad \mathrm{T}=0.030 \; .
\label{eq:param7}
\end{equation}
This parameter set is chosen to reproduce the \lamb{7}{Li} spectrum
which it does quite well, as can be seen by comparing the energies in
the last column of Table~\ref{tab:L7Li} with the experimental energies
at the left of Fig.~\ref{fig:lli7}. Note that an increase in one or both
of the `small' parameters S$_\vlam$ and T could reduce the excited-state
doublet splitting to the experimental value of 471\,keV. Also that these
two parameters have to take small values if they have the same signs
as in (\ref{eq:param7}). Tighter constraints on these parameters
come from the spectra of heavier p-shell hypernuclei (see later).

  Returning to the LS limit, the coefficient of $\vdel$ for the
ground-state doublet, derived from (\ref{eq:spin}), is 3/2. A similar
evaluation using the LS structure of the members of the excited-state
doublet gives 7/6 for the coefficient of $\vdel$. In this case, the
full expression for the splitting of the excited-state doublet is
\cite{dg78}
\begin{equation}
\Delta E = \frac{7}{6}\vdel + \frac{7}{3}\mathrm{S}_\vlam
 - \frac{14}{5}\mathrm{T} \; .
\label{eq:75doublet}
\end{equation}
This expression can be derived in a variety of ways using the
results in Appendix~\ref{sec:racsu2} or Appendix~\ref{sec:twobod}
but perhaps most easily by multiplying the coefficients of $\vdel$,
$S_\vlam$ and T for the $7/2^+$ state in Table~\ref{tab:L7Liabs},
for which twice the $2^-$ two-body matrix element [(\ref{eq:lstwo})
or (\ref{eq:jjtwo})] enters because the angular momenta are stretched
for the $7/2^+$ state, by 7/3 because their contribution measures the
shift from the centroid $2\,\overline{V} + \mathrm{S}_\mathrm{N}$ of the
$7/2^+$ and $5/2^+$ levels.

 The $\langle p_\mathrm{N}s_\vlam |V|p_\mathrm{N}s_\vsig\rangle$ matrix
elements were calculated from the YNG interaction SC97f(S) of \cite{akaishi00}
using harmonic oscillator wave functions with $b = 1.7$\,fm. These
matrix elements were multiplied by 0.9 to simulate the $\vlam$--$\vsig$
coupling of SC97e(S) and thus the observed doublet splitting for
\lamb{4}{He} (see \cite{akaishi00}). In the same parametrization as
for the $\vlam$N interaction
\begin{equation}
\overline\mathrm{V}'=1.45\quad \vdel'= 3.04\quad \mathrm{S}_\vlam' =
-0.085\quad \mathrm{S}_\mathrm{N}'  = -0.085 \quad \mathrm{T}'=0.157 \; .
\label{eq:lsparam7}
\end{equation}

 The YNG interaction has non-central components
but the dominant feature is a strong central interaction in the
$^3$S channel reflecting the second-order effect of the strong
tensor interaction in the $\vlam$N--$\vsig$N coupling. Because
the relative wave function for a nucleon in a p orbit and a
hyperon in an s orbit is roughly half s state and half p state,
the matrix elements coupling $\vlam$-hypernuclear and
$\vsig$-hypernuclear configurations are roughly a factor of two
smaller than those for the $\mathrm{A}\!=4\!$ system. Because the energy shifts
for the $\vlam$-hypernuclear states are given by $v^2/\Delta E$,
where $v$ is the coupling matrix element and $\Delta E\sim 80$ MeV,
the shifts in p-shell hypernuclei will be roughly a quarter of those
for $\mathrm{A}\!=\!4$ in favorable cases; e.g. 150 keV if the
$\vlam$--$\vsig$ coupling accounts for about half of the $\mathrm{A}\!=\!4$
splitting. For $\mathrm{T}\!=\!0$ hypernuclei, the effect will be
smaller because the requirement of a $\mathrm{T}\!=\!1$ nuclear core
for the $\vsig$-hypernuclear configurations
brings in some recoupling factors which are less than unity.
For example, in the case of the \lamb{7}{Li} ground state
\begin{eqnarray}
\lefteqn{\langle p^2(L\!=\!0\,S\!=\!1\,T\!=\!0)\,s_\varLambda,J=\!\frac{1}{2}\,
|\,V\,|\,p^2(L\!=\!0\,S\!=\!0\,T\!=\!1)\,s_\varSigma,J=\!\frac{1}{2}\rangle} \nonumber\\
& & =  2\sum_{\bar{S}} U(1/2\,1/2\,1/2\,1/2,1\bar{S})
U(1/2\,1/2\,1/2\,1/2,0\bar{S})
\langle ps_\varLambda,\bar{S}|V|ps_\varSigma,\bar{S}\rangle \nonumber\\
 & & = \frac{\sqrt{3}}{2}\, ({^3V} - {^1V}) = \frac{\sqrt{3}}{2}
\, \vdel'  \; .
\label{eq:7ligs}
\end{eqnarray}
Putting in the value for $\vdel'$ from (\ref{eq:lsparam7}) and taking
the actual value for $\Delta E$ of $\sim 88.5$\,MeV gives 78 keV for the
energy shift (cf. Table~\ref{tab:L7Liabs}). The result depends only on
$\vdel'$ because $\overline\mathrm{V}'$ connects only states with the same
core and because the spin-spin term is required to connect the core states
in (\ref{eq:7ligs}). In fact, the coefficients of the $\vlam$--$\vsig$
coupling parameters depend on isovector one-body density-matrix
elements connecting the core states (essentially $\langle\sigma\tau\rangle$
for $\vdel'$).

 A comparison of the ground-state doublet splitting for \lamb{7}{Li}
with that for \lamb{4}{He} (and \lamb{4}{H}) using modern YN interactions
with the same Monte Carlo~\cite{carlson90,carlson98}, or other few-body,
methods for both should provide a tight constraint on the strength of the
$\vlam$--$\vsig$ coupling because the contributions to the doublet
spacings are very different in the two cases.

\section{The p-shell Nuclei}
\label{sec:pshell}

 The structure of the spin-dependent operators in (\ref{eq:vlam})
means that their effects are most easily seen in an LS coupling basis.
In particular, either $\mathrm{L}\! =\! 0$ or $\mathrm{S}\! =\! 0$ for the core
isolates one of the parameters ($\vdel$ or S$_\vlam$, respectively).
The example of
\lamb{7}{Li} illustrated this point and is also a case in which
calculations can be made by hand. In the general case, a shell-model
calculation for the p shell is made with a phenomenological interaction
fitted to p-shell data. The wave functions are expanded on a
basis set (maximum dimension 14)
\begin{equation}
 H\Psi = E\Psi\quad\quad\quad  \Psi = \sum_i a_i \Phi_i \; ,
\label{eq:pham}
\end{equation}
where $\Phi$ could be expressed in $jj$ coupling
\begin{equation}
\Phi = |p_{3/2}^m (J_1T_1)p_{1/2}^{n-m} (J_2T_2); JT\rangle
\label{eq:jjbasis}
\end{equation}
or LS coupling (the Wigner supermultiplet scheme)
\begin{equation}
\Phi = |p^n[f]KLSJT\rangle \; .
\label{eq:lsbasis}
\end{equation}

 The purpose of the present section is to illustrate the structure
of p-shell nuclei in terms of the latter basis. This basis turns out
to be very good in the sense that the wave functions for
low-energy states are frequently dominated by one basis state, or
just a few basis states. This aids in the physical interpretation
of the structure. From the hypernuclear point of view, the contributions
of $\vdel$ and S$_\vlam$ depend only on the intensities of L and S
in the total wave function and these can be obtained in the weak-coupling
limit from a knowledge of the core wave function in an LS basis.

 $H$ is defined by two single-particle energies and 15 two-body
matrix elements. In terms of the relative coordinates of a pair
of nucleons, s, p and d states are possible for two p-shell
nucleons. There are 6 central matrix elements (one in each relative
state for $\mathrm{S} = 0,\,1$) which are attractive in spatially even states
and repulsive in odd states. The central part of the Millener--Kurath
interaction~\cite{mk75} (a single-range Yukawa interaction with
$b/\mu \!=\! 1.18$, potential strengths in MeV) illustrates this point
(the superscripts are $2T+1$ and $2S+1$).
\begin{equation}
 V^{11} = 32.0\quad\quad V^{31} = -26.88\quad\quad V^{13} = -44.8
\quad\quad V^{33} = 12.8
\label{eq:mk}
\end{equation}
There are 6 vector matrix elements,
2 arising from spin-orbit interactions in triplet p and d states
and 4 from antisymmetric spin-orbit (ALS) interactions that connect
two-body states with $\mathrm{S}\! = \!0$ and $\mathrm{S}\! =\! 1$ (these
are not present in the free interaction for identical baryons).
Finally, there are 3 tensor matrix elements in triplet p and d
states and connecting triplet s and d, states.

 The above approach is exemplified by the classic Cohen and
Kurath~\cite{ck65} fits to p-shell energy levels in terms of a constant
set of single-particle energies and two-body matrix elements. The
assumption of an A-independent interaction is a reasonable one for the
p shell because the rms charge radii of p-shell nuclei are rather constant,
as shown in Table~\ref{tab:rms}. This is basically because the p-shell
nucleons become more bound as more particles are added to the shell
and the rms radii of the individual orbits tend to stabilize
as nucleons are added.

\begin{table}
\centering
\caption{Root-mean-square charge radii (fm) of stable p-shell nuclei}
\label{tab:rms}
\begin{tabular}{cccccccccc}
\hline\noalign{\smallskip}
$^6$Li & $^7$Li & $^9$Be & $^{10}$B & $^{11}$B & $^{12}$C & $^{13}$C &
$^{14}$C & $^{14}$N & $^{15}$N \\
\noalign{\smallskip}\hline\noalign{\smallskip}
 2.57 & ~~2.41 & ~~2.52 & ~~2.45 & ~~2.42 & ~~2.47 & ~~2.44 & ~~2.56 &
~~2.52 & ~~2.59 \\
\noalign{\smallskip}\hline
\end{tabular}
\vspace*{-2mm}
\end{table}

 Cohen and Kurath obtained three different interactions by fitting
binding energies relative to $^4$He after the removal of an estimate
for the Coulomb energy. These interactions were designated as (8--16)2BME,
(6--16)2BME, and (8--16)POT where the mass ranges fitted are specified
and POT means that the 4 ALS matrix elements out of the 17 parameters
were set to zero. In the course of the hypernuclear studies described
here (and for other reasons) many fits have been made to a modern data
base of p-shell energy-level data. Only well determined linear combinations
of parameters (considerably less than 17) defined by diagonalizing
$\partial {\chi^2}/\partial {x_i}\partial {x_j}$ have been varied where
${\chi^2}$ measures the deviation of the theoretical and experimental
energies in the usual way and the $x_i$ are the parameters. The
fit69 and fit5 interactions in Table~\ref{tab:6liwfn} are examples
fitted to data on the $\mathrm{A}\!=\!6-9$ nuclei with only the central
and one-body interactions varied for fit69 and with the tensor
interaction and the one-body spin-orbit splitting fixed for fit5.

\begin{table}[b!]
\centering
\vspace*{-2mm}
\caption{Quantum numbers for p-shell nuclei}
\label{tab:u3u4}
\begin{tabular}{crccc}
\hline\noalign{\smallskip}
 U3 & SU3 & L & U4 & (TS)  \\
\noalign{\smallskip}\hline\noalign{\smallskip}
 $[2]$  &  ~~$(2\,0)$ & 0,2 & [11] & (01)(10) \\
 $[11]$ &  $(0\,1)$ & 1 & [2]  & (00)(11) \\
 $[3]$  &  $(3\,0)$ & 1,3 & [111] & $(\frac{1}{2}\,\frac{1}{2})$ \vspace{1.2pt}\\
 $[21]$ &  $(1\,1)$ & 1,2 & [21]  & $(\frac{1}{2}\,\frac{1}{2})
(\frac{1}{2}\,\frac{3}{2})(\frac{3}{2}\,\frac{1}{2})$ \vspace{1.2pt}\\
 $[111]$ &  $(0\,0)$ & 0 & [3] &  $(\frac{1}{2}\,\frac{1}{2})
(\frac{3}{2}\,\frac{3}{2})$ \\
 $[4]$  &  $(4\,0)$ & 0,2,4 & ~~[1111] & (00) \\
 $[31]$ &  $(2\,1)$ & 1,2,3 & [211]  & (01)(10)(11) \\
 $[22]$ &  $(0\,2)$ & 0,2 & [22] &   (00)(11)(02)(20) \\
 $[211]$ & $(1\,0)$ & 1  & [31] &   (01)(10)(11)(12)(21) \\
 $[41]$  &  $(3\,1)$ & 1,2,3,4 & [1] & $(\frac{1}{2}\,\frac{1}{2})$ \vspace{1.2pt}\\
 $[32]$ &  $(1\,2)$ & 1,2,3 & [221]  & $(\frac{1}{2}\,\frac{1}{2})
(\frac{1}{2}\,\frac{3}{2})(\frac{3}{2}\,\frac{1}{2})$ \vspace{1.2pt}\\
 $[311]$ &  $(2\,0)$ & 0,2 & [311] & $(\frac{1}{2}\,\frac{1}{2})
(\frac{1}{2}\,\frac{3}{2})(\frac{3}{2}\,\frac{1}{2})
(\frac{3}{2}\,\frac{3}{2})$ \vspace{1.2pt}\\
 $[221]$ &  $(0\,1)$ & 1 & [32] &  ~~$(\frac{1}{2}\,\frac{1}{2})
(\frac{1}{2}\,\frac{3}{2})(\frac{3}{2}\,\frac{1}{2})
(\frac{3}{2}\,\frac{3}{2})(\frac{1}{2}\,\frac{5}{2})
(\frac{5}{2}\,\frac{1}{2})$ \vspace{1.2pt}\\
 $[42]$  &  $(2\,2)$ & ~~0,$2^2$,3,4 & [11] & (01)(10) \\
 $[411]$ &  $(3\,0)$ & 1,3 & [2]  & (00)(11) \\
 $[33]$ &  $(0\,3)$ & 1,3 & [222]  & (00)(11) \vspace{1.pt}\\
 $[321]$ &  $(1\,1)$ & 1,2 & [321] & (01)(10)$(11)^2$(02)(20)(12)(21) \\
 $[222]$ & $(0\,0)$ & 0  & [33] &   (01)(10)(12)(21)(03)(30) \\
\noalign{\smallskip}\hline
\end{tabular}
\end{table}

In the supermultiplet basis, $[f]$KL label representations of
${\rm SU3}\supset {\rm R3}$ in the orbital space and $[\widetilde{f}]\beta$TS
label representations of ${\rm SU4}\supset {\rm SU2}\times {\rm SU2}$ in
the spin-isospin
space. Actually, $[f] =  [f_1 f_2 f_3]$ labels representations of U3
with $f_1\ge f_2\ge f_3$ and $f_1+ f_2+ f_3 = n$ and can be represented
pictorially by a Young diagram with $f_1$ boxes in the first row, $f_2$
in the second and $f_3$ in the third.
For a totally antisymmetric wave function,  $[\widetilde{f}]$ must be
the conjugate of $[f]$ and is obtained by interchanging the rows and columns
of the Young diagram. The length of the rows is then restricted to four.
In an oscillator basis, there is one quantum per particle in the p-shell
and $[f]$ labels also the symmetries of the quanta and the wave functions
have an SU3 symmetry labelled by $(\lambda\,\mu) = (f_1- f_2\, f_2 -f_3)$
with K and L given by
\begin{eqnarray}
 K & = & \mu,\,\mu-2,\,...,\, 1\ {\rm or}\ 0 \nonumber \\
 L & = & K,\,K+1,\,...,\, K+
\lambda \nonumber \\
 L & = & \lambda,\,\lambda-2,\,...,\, 1\ {\rm or}\ 0\ {\rm if}
\ K=0 \; .
\label{eq:su3qn}
\end{eqnarray}
For convenience, the allowed quantum numbers for $p^n$ configurations are
given in Table~\ref{tab:u3u4}. For $12-n$ particles, the L and TS
quantum numbers are the same and $(\lambda\,\mu)\to (\mu\,\lambda)$.

 In the following subsections, the spectra of selected p-shell nuclei for
$6 \leq \mathrm{A} \leq 14$ are presented and discussed in relation to the
supermultiplet structure of their p-shell wave functions (see the
tabulations covering the energy levels of light nuclei~\cite{tunl}
for more experimental information). Many of these nuclei form the
nuclear cores of hypernuclei discussed in detail in later sections.

\subsection{The Central Interaction}
\label{sec:central}

 The central interaction gives the bulk of the binding energy in p-shell
nuclei. It turns out to be essentially diagonal in the supermultiplet
basis and can be represented by 5 SU4 invariants.
\begin{equation}
 H  =  1.56\, n -  1.79 \sum_{i<j}I_{ij}
 -3.91 \sum_{i<j}P_{ij} +0.59\, \vec{L}^2 -1.08\, \vec{S}^2 + 0.59\, \vec{T}^2
\label{eq:hsu4}
\end{equation}
Here, the term linear in $n$ includes the centroid energy of the $p_{3/2}$
and $p_{1/2}$ orbits at $\mathrm{A}\!=\!5$ and takes care of the (constant)
one-body terms that arise from $\vec{L}^2$, $\vec{S}^2$, and $\vec{T}^2$.
The two-body identity operator counts the number of pairs $n(n-1)/2$
and the space-exchange operator counts the difference between the
numbers of spatially symmetric and antisymmetric pairs
$n_\mathrm{s} -n_\mathrm{a}$ given by
\begin{equation}
\langle \,[f]\,|\sum_{i<j}P_{ij}|\,[f]\, \rangle = \frac{1}{2}\sum_i f_i(f_i-2i+1)\; .
\label{eq:nsna}
\end{equation}
A rule of thumb is that this can be read off the Young diagram by
summing the number of pairs for each row and subtracting the number of
pairs for each column.
The relationship of the space-exchange operator to the quadratic Casimir
operator for SU4 is also worth noting,
\begin{equation}
 \sum_{i<j}P_{ij}  = 2\!n -\frac{1}{8}\!n^2 - \frac{1}{2}\!\mathrm{C(SU4)}\; ,
\label{eq:pij}
\end{equation}
where
\begin{equation}
\mathrm{C(SU4)} = \frac{1}{2} \sum_{i<j} \vec{\sigma}_i\cdot\vec{\sigma}_j
\,\vec{\tau}_i\cdot\vec{\tau}_j   + \vec{S}^2 + \vec{T}^2 +\frac{9}{4}n \; .
\label{eq:csu4}
\end{equation}
\begin{figure}[t]
\centering
\includegraphics[width=10.8cm]{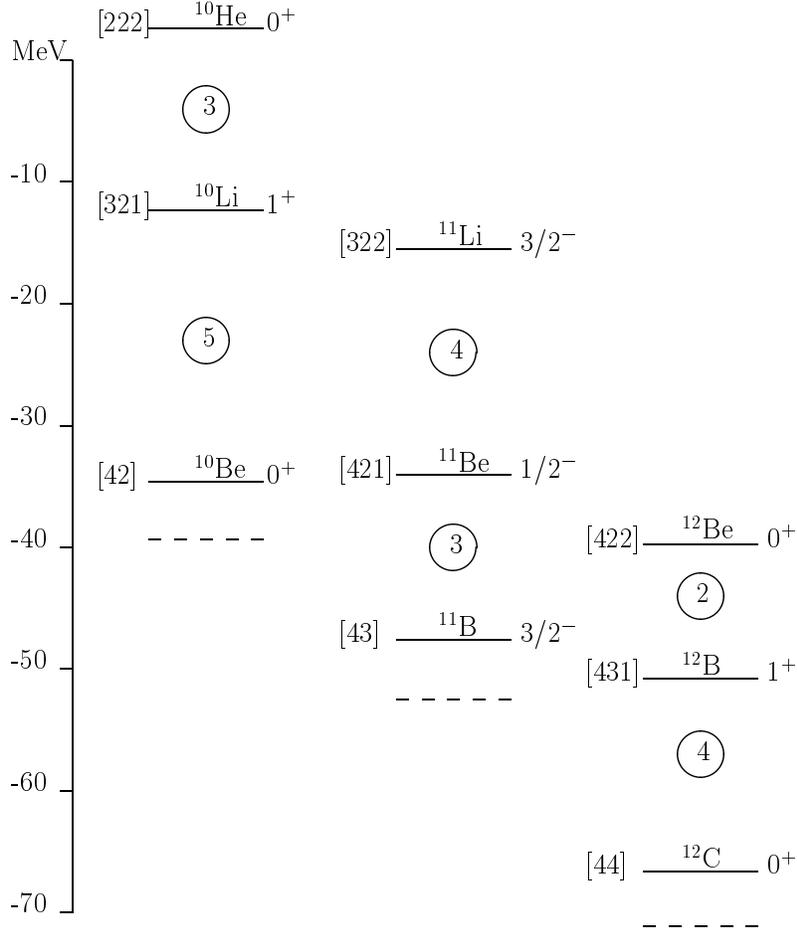}
\caption{Binding energy contributions from the SU4 invariant part
of the central interaction given by (\ref{eq:hsu4}). The dashed
lines give the result for the full p-shell Hamiltonian for the
ground states of the nuclei with the highest spatial symmetries.
The circled numbers are differences in $n_\mathrm{s}-n_\mathrm{a}$
for successive symmetries} \label{fig:hsu4}
\end{figure}

 Since there are only 6 independent central matrix elements, there
are only 6 independent operators to represent them. The remaining one
would connect the space and spin-isospin spaces, e.g. $\vec{l}_i\cdot
\vec{l}_j\,\vec{\tau}_i\cdot\vec{\tau}_j$. All the operators
in (\ref{eq:hsu4}) are SU3 scalars except for $\vec{L}^2$ which transforms
as a mixture of $(0\,0)$ and $(2\,2)$ tensors. An SU3 tensor expansion of
the central interaction contains four scalars and two $(2\,2)$ tensors,
so that the remaining operator has to have a $(2\,2)$ part to it.
If the coefficient of this extra operator is zero, the Hamiltonian in
(\ref{eq:hsu4}) represents the entire effect of the central interaction
throughout the shell (or for the range of nuclei fitted).

 The decomposition in (\ref{eq:hsu4}) comes from a fit to the
$\mathrm{A}\!=\!10-12$
nuclei and the coefficient of the extra $(2\,2)$ tensor is very small.
Figure~\ref{fig:hsu4} shows the binding energies given by (\ref{eq:hsu4}).
The dashed lines show the energies for the full Hamiltonian in the
cases of $^{10}$Be, $^{11}$B, and $^{12}$C. The gain is about 4\,MeV
in each case and is mostly due to turning on the spin-orbit interaction.
The circled numbers give the differences in the expectation value
of the space-exchange component ($n_s-n_a$) for successive spatial
symmetries. Four times the coefficient of the space-exchange operator
in (\ref{eq:hsu4}) is $\sim 15.6$\,MeV which is very close to the energy
of the first $\mathrm{T}\! = \!1$ states in $^{12}$C. Thus, it is evident
that the
SU4 invariant part of the central interaction gives a rather good account
of the general structure of p-shell nuclei. The spin-orbit interaction,
which transforms as $(1\,1)$ mixes spatial symmetries and L values. As
will be seen, the interplay between the spin-orbit and tensor
interactions can be very important and is quite subtle.

\subsection{Structure of $^6$Li, $^7$Li, and $^8$Be}
\label{sec:678}

\begin{figure}
\centering
\includegraphics[width=11.6cm]{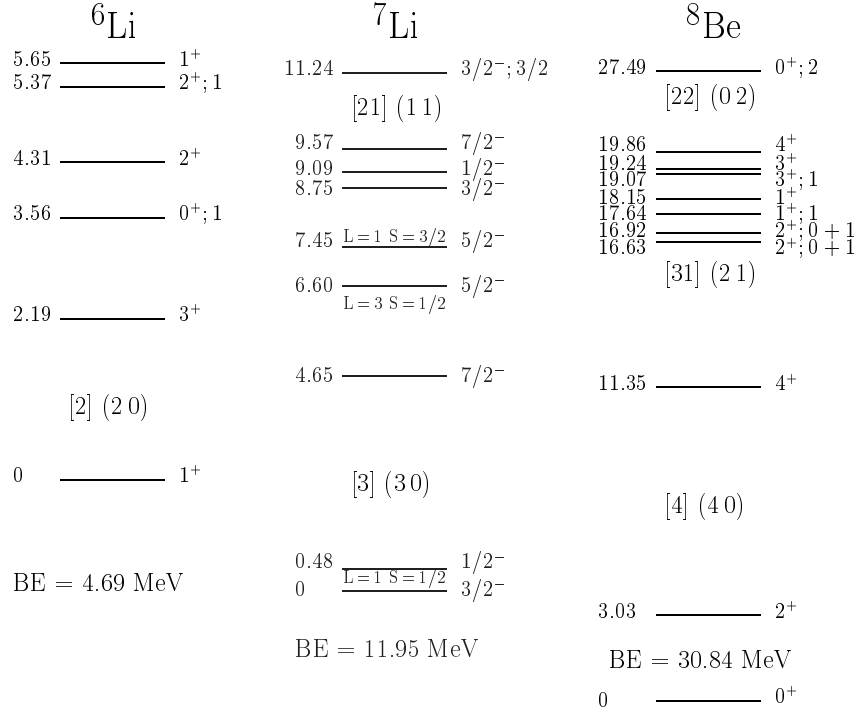}
\caption{Energy-level schemes for $^6$Li, $^7$Li, and $^8$Be with
the dominant spatial symmetry $[f]$, equivalently $(\lambda,\,\mu)$,
indicated for groups of levels and L and S for particular levels.
All energies are in MeV}
\label{fig:678}
\end{figure}

  The energy level schemes of these nuclei are shown in Fig.~\ref{fig:678}.
All the levels shown can be accounted for by p-shell calculations.
The lowest levels ($\mathrm{T}\! =\! 0$ for $^6$Li) have  well-developed
$\ualpha +{\rm d}$, $\ualpha +{\rm t}$, and $\ualpha +\ualpha$ cluster
structures. For harmonic
oscillator radial wave functions, coordinate transformations can be made
on the states with maximal spatial symmetry so that all the quanta
associated with the p-shell orbits reside on the  relative coordinate
between clusters formed from internal $0s$ wave functions (and the
center of mass is in a $0s$ state). These states must transform as
$(\lambda\,0)$ where $\lambda$ is the number of quanta. Oscillator
shell-model configurations beyond the p-shell are required to improve
the radial behavior of the relative wave functions.

 Wave functions for $^6$Li have been given in Table~\ref{tab:6liwfn}.
It can be seen that LS coupling is rather good and that the
$3^+;0$, $2^+;0$ and $1^+_2;0$ states form a triplet with
$\mathrm{L} \!=\! 2$, and $\mathrm{S} \!=\! 1$. Both vector and tensor
forces can contribute to the splitting of this triplet. The most
natural explanation is that the splitting is mainly due to the
one-body spin-orbit interaction, partly because the even-state
spin-orbit interaction acts in relative d states and the matrix
elements are small in a G-matrix derived from a realistic NN
interaction. This is not necessarily the case for a fitted
interaction. For example, the Cohen and Kurath interactions have
small one-body spin-orbit terms and substantial even-state and
antisymmetric spin-orbit terms that act in part like a one-body
spin-orbit interaction with a strength that depends linearly on
$n$ and ensures that the p-hole states at $\mathrm{A}\!=\!15$ are
split by just over 6\,MeV. The small quadrupole moment of $^6$Li
(experimentally $-0.082$\,fm$^2$) provides a constraint on the
balance of spin-orbit and tensor interactions. Writing
\begin{equation}
 |{^6{\rm Li}}(gs)\rangle = \alpha\,{^3S_1} +\beta\,{^3D_1}
+\gamma\,{^1P_1}
\label{eq:6ligs}
\end{equation}
leads to
\begin{equation}
  Q(^6\mathrm{Li}) = e^0 b^2 \left( \frac{4}{\sqrt{5}}\;\alpha\beta + \gamma^2
 -\frac{7}{10}\;\beta^2\right) \; ,
\label{eq:6liq}
\end{equation}
where $b\sim 1.7$\,fm is the oscillator parameter and $e^0\sim 0.815$
is the isoscalar effective charge
$(1+\delta e_\mathrm{p} + \delta e_\mathrm{n})/2$.
The direct tensor interaction coupling the $^3$S and $^3$D states gives
$\beta < 0$ while indirect coupling through the $^1$P state via the
spin-orbit interaction gives $\beta >0$. Putting in numbers from
Table~\ref{tab:6liwfn} shows that a small negative value for $\beta$ is
required. The B(M1$; 2^+;1\to 1^+;0) = (8.3\pm 1.5)\times 10^{-2}$\,W.u.
puts a similar restriction on $\beta$; briefly, the orbital
contribution connecting $^1$D to $^1$P and the spin contribution
connecting $^3$P to $^1$P are of the same sign while a spin contribution
of the opposite sign from $^1$D to $^3$D cannot be too large if the
B(M1) is to be reproduced. The interplay of spin-orbit and tensor
interactions in leading to small but important wave function admixtures
is a common feature in p-shell nuclei, most famously in the case of
the very hindered $^{14}$C$(\ubeta^-)$ decay (see later).

  In $^7$Li, the lowest four states form the ground-state band and have
$> 93$\% purity of the indicated LS configurations. The first
$\mathrm{T}\! =\! 3/2$ has a similar purity of [2\,1] symmetry with
$\mathrm{L}\! = \!1$ and $\mathrm{S}\! =\! 1/2$.
An interesting point is that the second $5/2^-$ state has a small
width for decay into the $\ualpha +{\rm t}$ channel despite its proximity
to the first $5/2^-$ level which has a large decay width into this
channel~\cite{tunl}. This means that the mixing matrix element between
the [3] and lowest [21] symmetry $5/2^-$ states has to be small. Only
a tensor interaction can connect the dominant components shown in
Fig.~\ref{fig:678} and this has to largely cancel with the spin-orbit
contribution arising from a modest [21] $^2$D component in the second state.

 The hypernucleus \lamb{8}{Li} (also \lamb{8}{Be}) was frequently
observed in emulsion studies~\cite{davis86} and analysis of the
characteristic decay mode
\begin{equation}
 ^8_\vlam\mathrm{Li} \to \upi^- + {^4\mathrm{He}} + {^4\mathrm{He}}
\label{eq:l8li}
\end{equation}
established a ground-state spin-parity of $1^-$ and provided
information on the mixing of configurations based on the ground-state
and first-excited state of $^7$Li. The configuration mixing is larger
than usual because core states are close together and share the same L
value. Both the ground-state spin of \lamb{8}{Li} and the mixing
provide restrictions on the nature of the $\vlam$N effective
interaction. To be studied by $\ugamma$-ray spectroscopy, the
$\mathrm{A}\!=\!8$ hypernuclei have to be formed by particle emission
from a heavier hypernucleus.

 The lowest $0^+$ and $2^+$ states of $^8$Be form the core for
bound states of \lamb{9}{Be} (discussed in detail later). The
$0^+$ state is unbound by 92\,keV and has a width of $\sim 6$\,eV
while the $2^+$ state has a width of $\sim 1.5$\,MeV. In the p-shell
model, these states have very pure [4] symmetry with a few percent
of [31] symmetry with $\mathrm{S}\! =\!1$. Because the Gamow-Teller operator
cannot change spatial quantum numbers, it is these small admixtures
in the $2^+$ wave function that account for the $\ubeta$ decays of
$^8$Li and $^8$B. The near degeneracies of pairs of `[31]' states with
the same J$^\pi$, different isospin, and similar space-spin wave functions
lead to isospin mixing that is especially strong for the 16.63-MeV and
16.92-MeV $2^+$ levels. The $\mathrm{T}\! =\! 1$ analogs of these levels form
the basis for the ground-state doublets of \lamb{9}{Li} and \lamb{9}{B}.
The \lamb{9}{Li} hypernucleus has been studied recently  via the
$^9\mathrm{Be}\,(\mathrm{e},\mathrm{e'}\mathrm{K}^+)\,$\lamb{9}{Li}
reaction~\cite{hallA}.

 The ground-state binding energies of the nuclei in Fig.~\ref{fig:678}
increase rapidly with the number of particles because the Pauli principle
permits up to two neutrons and two protons to correlate strongly in
spatially even states and take advantage of the strong central interaction
in relative s states, as quantified in Sect.~\ref{sec:central}.

\subsection{Structure of $^9$Be and $^{11}$C}
\label{sec:911}

\begin{figure}
\centering
\includegraphics[width=11.8cm]{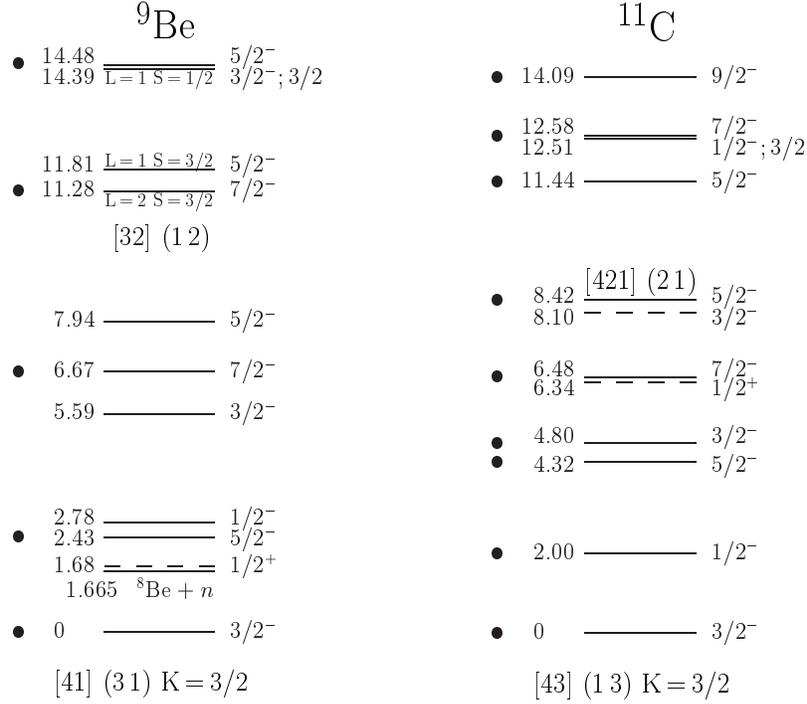}
\caption{Energy-level schemes for $^9$Be and $^{11}$C with
the dominant spatial symmetry $[f]$, equivalently $(\lambda,\,\mu)$,
indicated for groups of levels and L and S for particular levels.
All energies are in MeV. The bullets mark levels strongly populated in
proton knockout (or pickup) from $^{10}$B and the
$^{14}$N$(\mathrm{p},\ualpha)^{11}$C reaction}
\label{fig:911}
\end{figure}

 Partial energy level schemes of $^9$Be and $^{11}$C are given
in Fig.~\ref{fig:911}. These nuclei are paired together because,
with $p^5$ and $p^7$ configurations, they are related by a
particle-hole symmetry reflected in the conjugate SU3 representations.
Because the L values for the highest symmetry differ by steps of one
(see Table~\ref{tab:u3u4}), there are often two states with the same J
value. To some extent, these states can be organized into $\mathrm{K}\!=\!3/2$
and $\mathrm{K}\!=\!1/2$ bands. In fact, it was shown a long time
ago~\cite{kurath59}
that shell-model states for the p-shell nuclei have a large overlap with
states angular momentum projected from a Slater determinant made
from the lowest Nilsson model states (restricted to the p-shell
and with the same one-body spin-orbit interaction as in the shell-model
calculation) for some deformation. The deformations varied smoothly
with the number of nucleons and were prolate at the beginning of the
shell and oblate at the end of the shell. Thus, the lowest states of $^6$Li,
$^7$Li, and $^8$Be were obtained by filling the first $\mathrm{K}\!=\!1/2$
Nilsson orbit. The $\mathrm{K}\!=\!3/2$ orbit starts to fill at $^9$Be and the
second $\mathrm{K}\!=\!1/2$ orbit is relatively close in energy.
The ground-state of $^{11}$C (or $^{11}$B) would have three nucleons
(or a hole) in the $\mathrm{K}\!=\!3/2$ orbit.

 The connection to SU3 symmetry is quite close because
Elliott~\cite{elliott58} showed that all the angular momentum states
for a given SU3 representation could be projected out of a highest-weight
state characterized by numbers of quanta N$_z\! = a +\lambda+\mu$
and N$_\perp \!=2a+\mu$ with K$_\mathrm{L} \!= \mu, \mu -2,\ldots 1\
\mathrm{or}\ 0$
\begin{equation}
  |(\lambda\,\mu)K_L LM\rangle = \frac{1}{a(K_L L)}P^L_{MK}\,\Phi(HW) \; .
\label{eq:lproj}
\end{equation}
The highest-weight state is made up of asymptotic Nilsson orbits
(no spin-orbit interaction in this case). Something closer to
reality can be obtained by projecting from a product of the highest-weight
state and an intrinsic-spin wave function~\cite{elliott68}
\begin{eqnarray}
 |(\lambda\,\mu)K_J LSJM\rangle & = & N\, P^J_{MK}\, \Phi(HW)\chi(SK_S)
\nonumber \\
  &  = & \sum_L c(L)|(\lambda\, \mu)K_L LSJM\rangle \; ,
\label{eq:jproj}
\end{eqnarray}
so that a given state with good K in general contains a mixture
of L values. In SU3 codes~\cite{draayer73}, the basis of
(\ref{eq:lproj}) is used with the states orthogonalized with respect
to K$_\mathrm{L}$. The spin-orbit interaction can, and often does,
 mix L values to produce a good K$_\mathrm{J}$. It also mixes
$(\lambda\ \mu)$ and S values. For example, for the $^{11}$C ground state,
\begin{equation}
|(1\, 3)K\!=\!3/2\, J\!=\!3/2\rangle = \sqrt{21/26}\,|L\!=\!1\, S\!=\!1/2\rangle
- \sqrt{5/26}\,|L\!=\!2\, S\!=\!1/2\rangle \; .
\label{eq:c11gs}
\end{equation}
The CK816 interaction gives 0.7676 and $-0.4833$ for the coefficients,
meaning that $\mathrm{K}\!=\!3/2$ accounts for 81.3\% out of a total
82.3\% $[43]$ symmetry. There is 13.5\% $[421]$ symmetry in the wave function.

 An important point to notice is that the $^9$Be ground state is not
bound by much with respect to the neutron threshold (the $^9$B ground state
is unbound by 185\,keV with respect to proton emission). This is an effect
of the Pauli principle (embodied in the supermultiplet symmetry) which
strongly restricts the way in which an extra p-shell nucleon can interact
with a fully occupied orbit (in the Nilsson sense). On the other hand,
the $1s\,0d$ states, which are near zero binding at this mass number, can
couple to the $^8$Be core without restriction. In fact, the low-lying
positive-parity ($1\hbar\omega$) states also have a good SU3 symmetry,
namely $(6\,0)$ (typically $> 85$\%) obtained by coupling the $(2\,0)$
of the sd-shell nucleon to the $(4\,0)$ of the $^8$Be core.

 The levels of $^9$Be in Fig.~\ref{fig:911} marked by a bullet are
strongly excited in proton knockout, or pickup, from $^{10}$B~\cite{tunl}.
The strength is governed by a spectroscopic factor which, by definition, is
the square of the reduced matrix element of a creation operator connecting
the two states involved. The J dependence of the reduced matrix element
between basis states of the form (\ref{eq:lsbasis}) is contained in
a normalized $9j$ symbol via (\ref{eq:cprodsu2}). The reduced matrix element
that remains is just $\sqrt{n}$ times a one-particle coefficient of
fractional parentage (cfp) which defines how to construct a fully
antisymmetric $n$-particle state from antisymmetric
$(n-1)$-particle states coupled to the nth particle. Thus
\begin{eqnarray}
\lefteqn{\langle (\lambda\mu)\kappa LST||a^+||(\lambda'\mu')\kappa' L'S'T'
\rangle}\nonumber \\
& & = \sqrt{n}\sqrt{\frac{n_{f'}}{n_f}}\langle(\lambda'\mu')\kappa' L'
(1\ 0) 1 ||(\lambda\mu)\kappa L\rangle \langle [\widetilde{f'}]T'S'
[\widetilde{1}]1/2\,1/2 ||[\widetilde{f}]TS\rangle  \; ,
\label{eq:cfp}
\end{eqnarray}
where the Clebsch-Gordan coefficients for
$\mathrm{SU3}\supset\mathrm{R3}$~\cite{draayer73,jahn51}
and $\mathrm{SU4}\supset\mathrm{SU2}\times\mathrm{SU2}$~\cite{jahn51,hecht69}
result from
applications of the Wigner-Eckart theorem for SU3 and SU4 and the weight
factor $n_{f'}/n_f$ is the ratio of dimensions of representations
of the symmetric groups S$_{n-1}$ and S$_n$~\cite{jahn51}. The $[f']$ are
found by removing one box from the Young diagram for $[f]$ in all
allowed ways. Examples of the weight factors for $^9$Be and $^{10}$B are
given in Table~\ref{tab:weight}.

\begin{table}
\centering
\caption{Weight factors for $^9$Be and $^{10}$B}
\label{tab:weight}
\begin{tabular}{cccc}
\hline\noalign{\smallskip}
 $^{9}{\rm Be}\to {^{8}{\rm Be}}+n$ &
~~~$\sqrt{n_{f'}/n_f}$~~~ & $^{10}{\rm B}\to {^{9}{\rm Be}}+p$ &
~~~$\sqrt{n_{f'}/n_f}$~~~ \\
\noalign{\smallskip}\hline\noalign{\smallskip}
$[41]\rightarrow [4]$ & $\sqrt{\frac{1}{4}}$ & $[42]\rightarrow [41]$ &
 $\sqrt{\frac{4}{9}}$ \vspace{1.5pt}\\
$\phantom{[41]}\rightarrow [31]$ & $\sqrt{\frac{3}{4}}$ &
$\phantom{[42]}\rightarrow [32]$ & $\sqrt{\frac{5}{9}}$ \\
\noalign{\smallskip}\hline
\end{tabular}
\end{table}

 Because states with different supermultiplet symmetry are widely
separated, the one-particle removal strength is in general
complex~\cite{ck67}. The same is true for two-particle~\cite{ck70}
and three-particle~\cite{km75} removal but less so for the removal
of an $\ualpha$ particle~\cite{kurath73} because the removed $p^4$
configuration is an SU4 scalar. Pickup reactions provide a powerful
way of identifying predominantly p-shell states. Stripping reactions
are also very useful but can strongly populate states in which
particles reside in higher shells (usually the next shell).

 Table~\ref{tab:weight} shows that one reason why the binding energy
of $^9$Be is low with respect to $^8$Be is that 3/4 of the parentage
of the $^9$Be ground state goes to highly-excited states of $^8$Be
and $^8$Li. The weight factors for $^{10}$B show that the parentage
is almost equally divided between states of [41] and [32] symmetry.
In fact, the lowest three states seen strongly in knockout are mainly
[41] symmetry and the upper two states are mainly [32] symmetry.
The upper $7/2^-$ state has a spectroscopic strength that is a factor
of two larger than that for the lower $7/2^-$ state~\cite{tunl}. In
the pure symmetry limit, this factor is $\sim 7$. The mixing of the
two basis configurations needed to obtain the experimentally measured
ratio is small. This is another case in which the balance between
vector and tensor interactions in the mixing matrix element is
important and different p-shell interactions tend to give rather
different results for the ratio of strengths for the $7/2^-$ states.

 A final observation for $^9$Be is that the 11.81-MeV $5/2^-$ state is fed
very strongly in the $\ubeta^-$ decay of $^9$Li~\cite{tunl} because it
has largely the same spatial quantum numbers as the initial state.
In fact, the B(GT) value is much larger than one would expect, perhaps
because of difficulties in analysing the $\ualpha+\ualpha + n$ final
state. The analogous $\ubeta^+$ decay of $^9$C~\cite{tunl} has close to the
strength expected from shell-model calculations.

 As expected, the $^{11}$C ($^{11}$B) spectrum shows many similarities
to the $^9$Be spectrum. The positive-parity states are now more bound
with respect to the nucleon threshold and, indeed, $^{11}$Be
has a $1/2^+$ ground state 0.32\,MeV below the $1/2^-$ state.
Because two particles can be promoted to the sd shell without breaking
up the $^8$Be core, $(sd)^2$ states are found quite low in energy,
starting with the 8.10-MeV $3/2^-$ level. One-neutron removal from
$^{12}$C is limited to the first two $3/2^-$ states and the first
$1/2^-$ state ($[44]\to [43]$ is unique). However, triton removal from
$^{14}$N via the $^{14}$N$(\mathrm{p},\ualpha)^{11}$C reaction~\cite{maples71},
and aided by the $^3$D character of the $^{14}$N ground state, strongly
populates all the $\mathrm{T}\! =\! 1/2$ p-shell states included in
Fig.~\ref{fig:911} (for theory, see \cite{km75}).

\subsection{Structure of $^{10}$B and $^{10}$Be}
\label{sec:b10be10}

\begin{figure}[b]
\centering
\includegraphics[width=11.8cm]{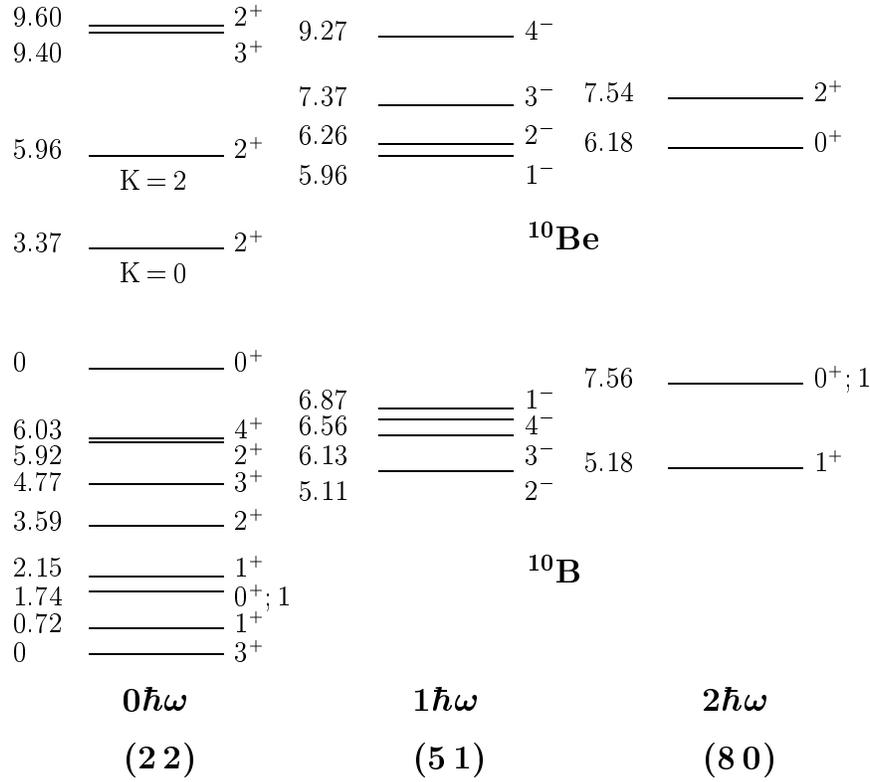}
\caption{Energy-level schemes for $^{10}$B (bottom) and $^{10}$Be
(top). All energies are in MeV. All states have mainly [42]
spatial symmetry except for the 9.60-MeV $2^+$ level of $^{10}$Be,
which has mainly [33] symmetry. The neutron and $\ualpha$
thresholds in $^{10}$Be are at 6.812\,MeV and 7.410\,MeV. The
$\ualpha$, deuteron, and proton thresholds in $^{10}$B are at
4.461\,MeV, 6.027\,MeV, and 6.586\,MeV} \label{fig:b10be10}
\end{figure}

 Energy-level schemes of $^{10}$B and $^{10}$Be are given in
Fig.~\ref{fig:b10be10}. All the negative-parity states are shown.
They are low in energy for the same reason that positive-parity states
come low in $^9$Be. Now, low-lying $(sd)^2$ states are possible
because two
p-shell nucleons that are strongly affected by the Pauli principle
can be promoted to the sd shell without breaking up the [4] symmetry
for the first four p-shell nucleons. Shell-model calculations show
that all the states (except one) have the highest spatial symmetry
and are dominated by the leading SU3 symmetries, as indicated in
Fig.~\ref{fig:b10be10}.

 The structure of the p-shell states of $^{10}$B is interesting and is
important  for hypernuclear physics because  $^{10}$B forms the core
for \lam{11}{B} which has been  studied with the Hyperball detector.
Six $\ugamma$ rays were observed but not all of them can be placed in a
decay scheme. Even for those that can be placed with reasonable certainty,
there are some puzzles (see later).

 The $(2\,2)$ representation of SU3 contains two $\mathrm{L}=2$ states
[see Table~\ref{tab:u3u4} or (\ref{eq:su3qn})] and this is the only
case in the p-shell for which the K$_\mathrm{L}$ quantum number is required.
For $^{10}$Be, $\mathrm{S}\!=\!0$ and the K assignments are clear and
understandable in terms of two particles in the $\mathrm{K}\!=\!3/2$
Nilsson orbit or one each in the $\mathrm{K}\!=\!3/2$ and $\mathrm{K}\!=\!1/2$
orbits (these orbits have K$_\mathrm{L}\!=\!1$ and K$_\mathrm{S}\!=\pm 1/2$).
For $^{10}$B, $\mathrm{S}\!=\!1$ and the K$_\mathrm{L}\!=\!0$ states are the
0.72-MeV $1^+$ state  with $\mathrm{L}\!=\!0$ and the $\mathrm{L}\!=\!2$ triplet
of states at 2.15, 3.59, and 4.77\,MeV. The ground state has $\mathrm{K}\!=\!3$,
and mostly $\mathrm{L}\!=\!2$, and is connected by a very strong E2 transition
to the $4^+$ level at 6.03\,MeV, there being a predicted but unobserved
$5^+$ level at higher energy. The 5.92-MeV $2^+$ level is mainly
$\mathrm{L}\!=\!2$ with K$_\mathrm{L}\!=\!2$ and in this sense is part of a
triplet involving the $3^+$ ground state and a $1^+$ configuration predicted
at higher energy. This triplet has the property of being very strongly split by the
spin-orbit interaction while the K$_\mathrm{L}\!=\!0$ triplet remains much more
compact. Electromagnetic transitions in $^{10}$B have been investigated in
great detail in the past~\cite{tunl} and it is from various selection
rules that the K quantum numbers can be assigned. In particular,
strong isovector M1 transitions must connect states with the same
K$_\mathrm{L}$.

\begin{figure}
\centering
\includegraphics[width=11.8cm]{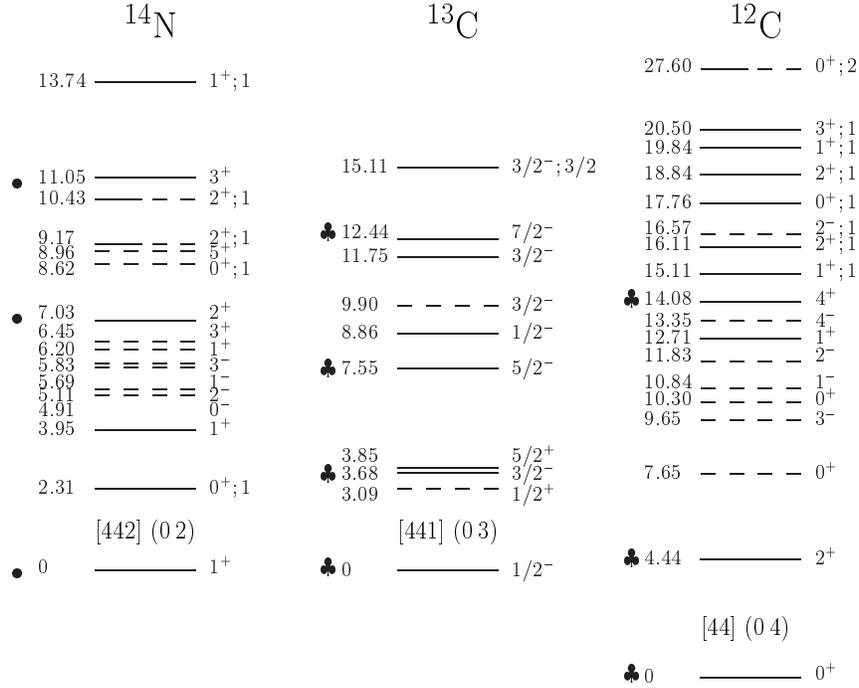}
\caption{Energy-level schemes for $^{12}$C, $^{13}$C, and $^{14}$N.
All energies are in MeV. For $^{12}$C and $^{13}$C, the club signs identify
the members of the ground-state bands with the dominant symmetry indicated.
For $^{14}$N, the bullets indicate a triplet of states with $^3$D two-hole
configurations. Dashed lines indicate non p-shell states.
The lowest particle thresholds are $\ualpha$ at
7.367\,MeV in $^{12}$C, neutron at 4.946\,MeV in $^{13}$C, and
proton at 7.551\,MeV in $^{14}$N}
\label{fig:1214}
\end{figure}

\subsection{Structure of $^{12}$C, $^{13}$C, and $^{14}$N}
\label{sec:1214}

 The energy level schemes of $^{12}$C, $^{13}$C, and $^{14}$N are
given in Fig.~\ref{fig:1214}. These nuclei are in a sense the
particle-hole conjugates of the nuclei shown in Fig.~\ref{fig:678}.
However, the effective spin-orbit interaction, indicated by the more
than 6\,MeV separation of the single-hole states of $^{15}$N and
$^{15}$O, is much larger. The larger spin-orbit interaction tends
to break the supermultiplet symmetry. Nevertheless, the content of
the highest symmetry in the ``ground-state'' bands is typically $>70$\%
and often higher.

There are now an increasing number of
``intruder'' levels marked by dashed lines. In $^{12}$C, they include
the Hoyle state at 7.65\,MeV which is certainly not accounted for
in shell-model calculations up to $2\hbar\omega$. The negative-parity
states are, however, quite well accounted for in $1\hbar\omega$
shell-model calculations and have dominantly [44] symmetry and
$(3\,3)$ SU3 symmetry. The $0^+;2$ state is known to have a large,
or even dominant, $(sd)^2$ component.

In $^{13}$C, the extra p-shell nucleon is not well bound with respect
to $^{12}$C, the neutron threshold being at 4.95\,MeV (cf. $^9$Be vs.
$^8$Be) and positive-parity states, again unhindered by the Pauli
principle, appear at low energies. The 8.86-MeV and 11.75-MeV levels
are the lowest states with the [432] symmetry of the 15.11-MeV $3/2^-$
state, while the 9.90-MeV level is the lowest $(sd)^2$~state.

 In $^{14}$N, the lowest member of the marked group  of predominantly
$^3$D two-hole states has become the ground state with the
3.95-MeV level being the predominantly $^3$S state. The ground
state is also predominantly two $p_{1/2}$ holes (there is an
overlap of $\sqrt{20/27}$ with the $^3$D configuration). The
structure of the $^{14}$N ground state is the important factor in
the slowness of the $^{14}$C $\ubeta^-$ decay which is hindered by
about six orders of magnitude compared with a strong allowed
decay. Consider the following wave functions for the initial and
final states in the the $\ubeta^-$ decay
\begin{eqnarray}
| {^{14}{\rm C}(0^+;1)}\rangle & = & ~~0.7729\ ^1S + 0.6346\ ^3P \nonumber \\
| {^{14}{\rm N}(1^+;0)}\rangle & = & -0.1139\ ^3S + 0.2405\ ^1P - 0.9639\ ^3D
\label{eq:14gs}
\end{eqnarray}
The Gamow-Teller matrix element is proportional to
\begin{equation}
\sqrt{3}a(^1S)\,a(^3S) + a(^1P)\,a(^3P)
\label{eq:14GT}
\end{equation}
and for the wave functions above the matrix element is $\simeq 0$.
This is because the tensor interaction, essentially
$\langle s|V_\mathrm{T}|d\rangle$,
was chosen to ensure the cancellation and kept fixed during a p-shell fit.
This is another case where the spin-orbit interaction alone gives the
wrong sign for the $^3$S amplitude and a tensor interaction gives the
opposite sign (see \cite{rose68} for the history). Keeping the tensor
interaction fixed leads to improvements in most of the cases for which
the balance of tensor and vector interactions is important.

 The above cancellation of the Gamow-Teller matrix element also plays an
important role in the analogous M1 transition in $^{14}$N. The absence
of the normally dominant spin contribution to an isovector M1 transition
leads to a rather small B(M1) dominated by the orbital contribution.
This turns out to be important for understanding the properties of
\lam{15}{N} which has been studied with the Hyperball.

 Finally, the intruder positive-parity levels of $^{14}$N shown in
Fig.~\ref{fig:1214} are of $(sd)^2$ character, as one would expect from
the presence of the low-energy positive-parity states in $^{13}$C and
$^{13}$N, and in analogy to the $\mathrm{A}\!=\!10$ nuclei. The $2^+;1$ levels
have long been known to be of strongly mixed p-shell and $(sd)^2$
character.

\section{The p-shell Hypernuclei}
\label{sec:phyp}

 The structure of \lamb{7}{Li} has already been discussed in
Sect.~\ref{sec:L7Li} because, as an introduction to p-shell
hypernuclei, it is a simple case with a $p^2$ $^6$Li core that is
amenable to hand calculation. This example was also used to compare
and contrast the effects of $\vlam$--$\vsig$ coupling in the
s-shell and p-shell hypernuclei.

 Following the survey of p-shell structure in terms of the LS-coupling
supermultiplet basis in Sect.~\ref{sec:pshell}, this section is devoted
to presenting the results obtained with the Hyperball on heavier
p-shell hypernuclei and giving interpretations in terms of the
underlying p-shell structure and effective YN interactions. The hypernuclei
for which results have been obtained with the Hyperball in experiments
at KEK and BNL are \lamb{7}{Li}, \lamb{9}{Be}, \lam{10}{B}, \lam{11}{B},
\lam{12}{C}, \lam{15}{N}, and \lam{16}{O}. For \lam{16}{O}, the calculation
is a particle-hole calculation and for \lam{15}{N}, the calculation is
similar to that for \lamb{7}{Li} in that there are two p-shell holes instead
of two p-shell particles.

\subsection{The Shell-Model Calculations}
\label{sec:sm}

 The Hamiltonian
 \begin{equation}
  H = H_\mathrm{N} + H_\mathrm{Y} + V_\mathrm{NY} \; ,
\label{eq:hamyn2}
\end{equation}
and the weak-coupling basis were introduced in (\ref{eq:hamyn})
and (\ref{eq:basis}). The formalism for the hypernuclear shell-model
calculations is presented in Sect.~3.1 of~\cite{auerbach83} but some
of the basic formulae are given here for completeness. The YN
interaction can be written in terms of products of two creation
and two annihilation operators with coefficients that are essentially
the two-body matrix elements. The $a^+a^+a\,a$ product can be recoupled
in any convenient order using any convenient coupling scheme. In the
present case, it is convenient to write the operator in terms of $a^+a$
pairs for the nucleons and hyperons so that we have a zero-coupled product
of operators for separate spaces for which the matrix elements may be
separated using the formulae in Appendix~\ref{sec:racsu2}. Formally,
\begin{equation}
 V  = \sum_\alpha C(\alpha) \left[\left[
a^{+}_{j_\mathrm{N}}\widetilde{a}^{ }_{j_\mathrm{N}'}
\right]^{J_\alpha T_\alpha}\left[ a^{+}_{j_\mathrm{Y}}
\widetilde{a}^{ }_{j_\mathrm{Y}'}\right]^{J_\alpha T_\alpha}\right]^{00} \; ,
\label{eq:vcross}
\end{equation}
where $\alpha$ stands for all the quantum numbers and the properly
phased annihilation operators are given by
\begin{equation}
a_{jm\frac{1}{2}m_t} =
(-)^{j-m+\frac{1}{2}-m_t}\ \widetilde{a}_{j-m\frac{1}{2}-m_t}\; ,
\label{eq:annih}
\end{equation}
and
\begin{eqnarray}
  C(\alpha) & = & \sum_{KT}
\left(\begin{array}{lll} j_\mathrm{N} & j_\mathrm{Y} & K \\
j_\mathrm{N}' & j_\mathrm{Y}' & K \\ J_\alpha &
J_\alpha & 0 \end{array} \right)
\left(\begin{array}{lll} 1/2 & t_\mathrm{Y} & T \\ 1/2 & t_\mathrm{Y}' & T \\
T_\alpha & T_\alpha & 0 \end{array} \right) \nonumber \\
 &   & ~\times
\widehat{K}\widehat{T} \, \langle j_\mathrm{N}j_\mathrm{Y}t_\mathrm{Y};KT\,|\,V\,|\,
j_\mathrm{N}'j_\mathrm{Y}'t_\mathrm{Y}';KT\rangle \; .
\label{eq:calpha}
\end{eqnarray}
Then
\begin{eqnarray}
\lefteqn{ \langle\alpha_\mathrm{c} J_\mathrm{c} T_\mathrm{c}, j_\mathrm{Y}
t_\mathrm{Y};JT\,|\,V_\mathrm{NY-NY'}|
 \alpha_\mathrm{c}' J_\mathrm{c}' T_\mathrm{c}', j_\mathrm{Y}' t_\mathrm{Y}';JT
\rangle }\nonumber \\
 & & =  \sum_\alpha  C(\alpha)
\left(\begin{array}{lll} J_\mathrm{c}' & J_\alpha & J_\mathrm{c} \\
j_\mathrm{Y}' & J_\alpha & j_\mathrm{Y} \\
J & 0 & J \end{array} \right)
\left(\begin{array}{lll} T_\mathrm{c}' & T_\alpha & T_\mathrm{c} \\
t_\mathrm{Y}' & T_\alpha & t_\mathrm{Y} \\
T & 0 & T  \end{array} \right) \nonumber\\
& & ~\times {\widehat{J_\alpha}\widehat{T_\alpha}
 \over \widehat{j_\mathrm{Y}}\widehat{t_\mathrm{Y}}}\,
\langle \alpha_\mathrm{c} J_\mathrm{c} T_\mathrm{c}||
(a^+_{j_\mathrm{N}}\tilde{a}^{ }_{j_\mathrm{N}'})^{J_\alpha T_\alpha}||
\alpha_\mathrm{c}' J_\mathrm{c}' T_\mathrm{c}' \rangle \; .
\label{eq:manybody}
\end{eqnarray}
The basic input from the p-shell calculation is thus a set of
one-body density-matrix elements between all pairs of nuclear
core states that are to be included in the hypernuclear shell-model
calculation. As noted in Sect.~\ref{sec:L7Li}, experimental energies
are used for the diagonal core energies where possible.

 The one-body transition density that governs the cross section
for the formation of a particular hypernuclear state is (see Sect.~3.2
of \cite{auerbach83})
\begin{eqnarray}
 \lefteqn{\langle p^{n-1}\alpha_\mathrm{c} J_\mathrm{c} T_f, j_\varLambda 0;J_fT_f||
\left(
a^+_{j_\varLambda}\widetilde{a}^{ }_{j_\mathrm{N}}\right)^{\Delta J 1/2}||
p^n\alpha_i J_i T_i \rangle } \nonumber \\
& &  = (-)^{j_\mathrm{N} +j_\varLambda -\Delta J}\,
U(J_ij_\mathrm{N}J_fj_\varLambda,J_\mathrm{c}\Delta J)
\langle p^{n-1}\alpha_\mathrm{c} J_\mathrm{c} T_f||\widetilde{a}_{j_\mathrm{N}}
||p^n\alpha_i J_i T_i
\rangle \; .
\label{eq:formation}
\end{eqnarray}
An important result is that in the weak-coupling limit the total
strength for forming the states in a weak-coupling multiplet
(summing over $J_fj_\vlam$) is proportional to the pickup spectroscopic
factor from the target \cite{auerbach83}. To see the consequences
of the spin-flip characteristics of the reaction used to produce the
hypernuclear states, it is useful to change the coupling from
$(j_\mathrm{N}j_\varLambda)\Delta J$ to $(l_\mathrm{N}l_\varLambda)\Delta L
\Delta S\Delta J$ using (\ref{eq:9jsu2a}).

\subsection{The \lamb{9}{Be} Hypernucleus}
\label{sec:lbe9}

\begin{figure}
\centering
\includegraphics[width=9.0cm]{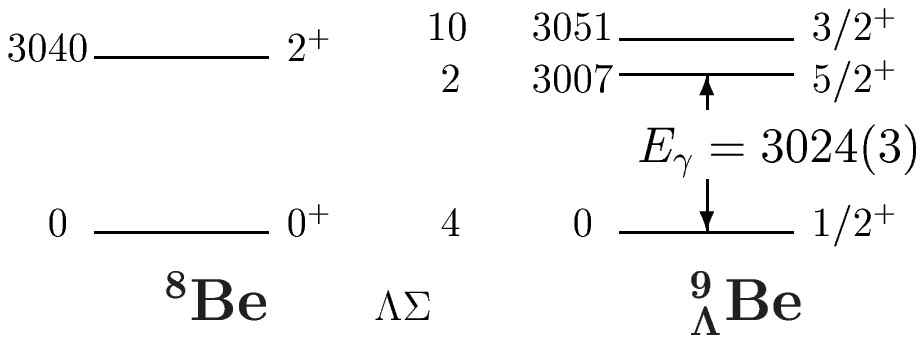}
\caption{Energy levels of $^9_\varLambda$Be and the $^8$Be core.
The small shifts due to $\vlam$--$\vsig$ coupling are shown in the
center. All energies are in keV. The measured $\ugamma$-ray
energies are 3024(3) and 3067(3) keV giving a doublet separation
of 43(5) keV \cite{tamura05}} \label{fig:lbe9}
\end{figure}

 The bound-state spectrum for $^9_\varLambda$Be is shown in Fig.~\ref{fig:lbe9},
which gives the $\ugamma$-ray energies from an analysis of the
BNL E930 data \cite{akikawa02,tamura05}, for the parameter set in
(\ref{eq:param7})
used for \lamb{7}{Li}. An earlier experiment with NaI detectors~\cite{may83}
observed a $\ugamma$ ray at 3079(40)\,keV and put an upper limit of
100\,keV on the doublet splitting.

 The breakdown of the doublet splitting is given in Table~\ref{tab:lbe9}.
In the LS limit for $^8$Be, the $2^+$ wave function has $\mathrm{L}\!=\!2$ and
$\mathrm{S}\!=\!0$. Then, only the coefficient of S$_\vlam$ survives and takes
the value $-5/2$ as can be seen from an equation analogous to
(\ref{eq:spin}) with $\vec{S}_\mathrm{c}$ replaced by $\vec{L}_\mathrm{c}$. In the
realistic case, the contributions of S$_\vlam$ and T work against
those from $\vdel$ and the $\vlam$--$\vsig$ coupling (small in this
case because the $\vsig$ has to be coupled to $\mathrm{T}\!=\!1$ states of the
core with a different symmetry from the $\mathrm{T}\!=\!0$ states). A similar
thing happens for the excited-state doublet of \lamb{7}{Li} and the
experimental results for both doublets restrict the combined effect
of S$_\vlam$ and T to be small.

\begin{table}[b]
\centering
\caption{Contributions from $\vlam$--$\vsig$ coupling and the
spin-dependent components of the effective $\vlam$N interaction
to the $3/2^+$, $5/2^+$ doublet spacing in \lamb{9}{Be}.
The spectrum is shown on the right hand side of Fig.~\ref{fig:lbe9}.
As in Table~\ref{tab:L7Li}, the first line gives the coefficient of each
parameter and the second line gives the actual energy contributions in keV}
\label{tab:lbe9}
\begin{tabular}{rrrrrr}
\hline\noalign{\smallskip}
 $\vlam\vsig$ & $\vdel$ & S$_\vlam$ & S$_\mathrm{N}$ &  T &
~$\Delta E$  \\
\noalign{\smallskip}\hline\noalign{\smallskip}
   & ~$-0.033$ & ~$-2.467$ & ~0.000 & ~0.940 & \\
 $-8$ &  $-14$  &  37  &  0  &   28  &  44  \\
\noalign{\smallskip}\hline
\end{tabular}
\end{table}

\begin{figure}
\centering
\includegraphics[width=10.0cm]{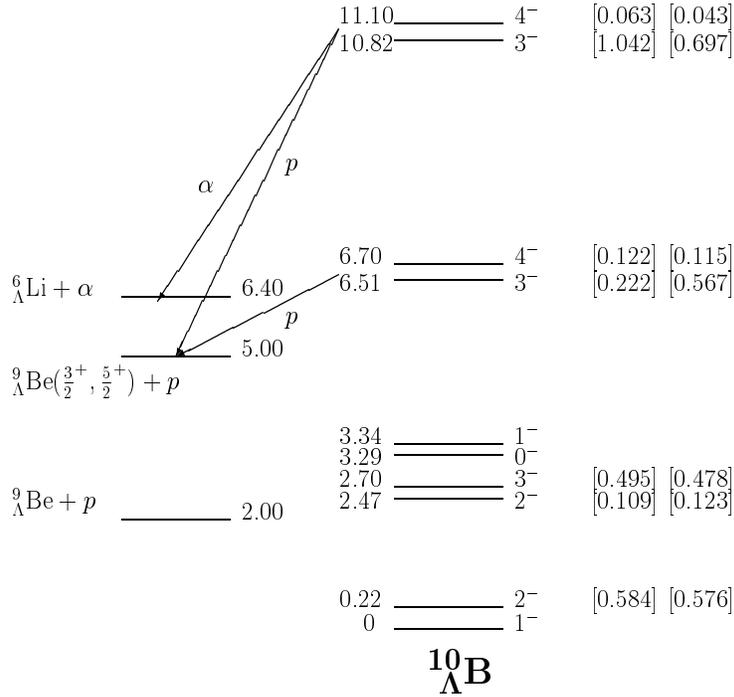}
\caption{Proton decay of \lam{10}{B} to \lamb{9}{Be}.
Formation strengths for non-spin flip production in the \Kpi
reaction are given on the right for two p-shell models. Thresholds
for particle decay of the \lam{10}{B} states are given on the
left. All energies are in MeV}
\label{fig:lbe9p}
\end{figure}

 The parameter set chosen puts the $3/2^+$ state above the $5/2^+$
state but the order is not determined by this experiment. However, in
the 2001 run of BNL E930 on a $^{10}$B target, only the upper level
is seen following proton emission from \lam{10}{B}. It can then be
deduced that the $3/2^+$ state is the upper member of the doublet
via the following reasoning. Four states of $^9$B are
strongly populated by neutron removal from $^{10}$B \cite{tunl}
and the hypernuclear doublets based on these states are shown in
Fig.~\ref{fig:lbe9p}. The structure factors which govern the
population of these states are given at the right of the figure for
two p-shell interactions. As discussed in Sect.~\ref{sec:911}, the
relative neutron pickup strength to the two $7/2^-$ states which
give rise to the $3^-$/$4^-$ doublets above the
$^9_\vlam{\rm Be}^* +\mathrm{p}$ threshold is very sensitive to the
non-central components of the p-shell interaction.
Formation of the $3^-$ states is favored for the
dominant $p_{3/2}$ removal by the coupling to get $\Delta L\! =\!1$
and $\Delta S \!=\!0$. The proton decay arises from $^9{\rm B}(7/2^-)\to
{^8\mathrm{Be}(2^+) + \mathrm{p}}$ in the core. The $4^-$ states proton decay
to $^9_\vlam{\rm Be}(5/2^+)$ and from the recoupling
$(2^+\times p_{3/2})7/2^-\times s_\vlam\to  (2^+\times s_\vlam)J_f
\times p_{3/2}$, governed by
\begin{equation}
(-)^{3/2+J_f-3}\ U(3/2\,2\,3\,1/2,7/2\,J_f) \; ,
\label{eq:9recoup}
\end{equation}
one finds that the $3^-$ states proton decay to the $3/2^+$ and $5/2^+$
states in the ratio of 32 to 3. Overall, the the $3/2^+$ state is
favored by a factor of more than 3. The only caveat to this argument
is that the uppermost $3^-$ state doesn't $\ualpha$ decay too much.

\subsection{The \lam{16}{O} Hypernucleus}
\label{sec:lo16}

\begin{figure}[b]
\centering
\includegraphics[width=11.0cm]{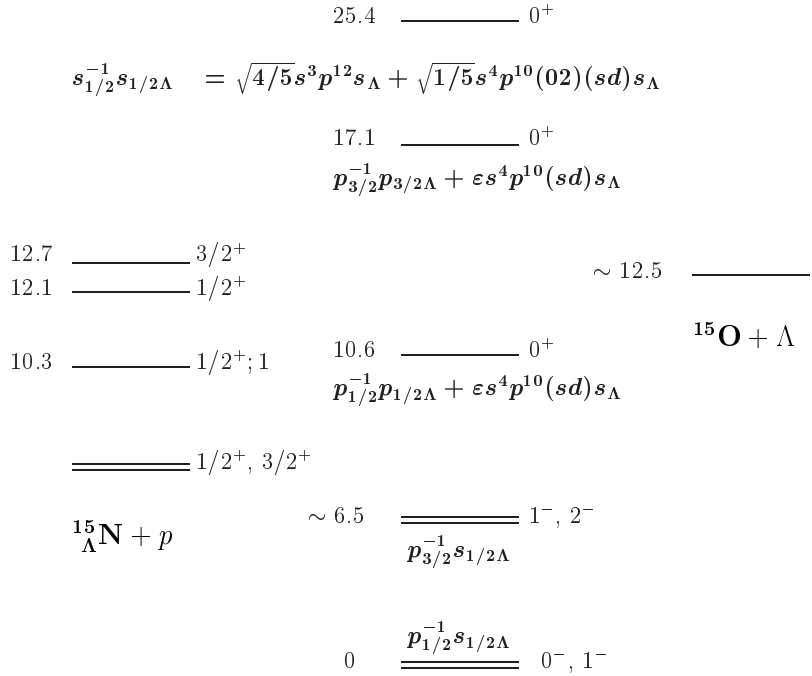}
\caption{The energies of $1^-$ and $0^+$ states of \lam{16}{O} that
are strongly populated in the $^{16}$O$(\mathrm{K}^-,\upi^-)$\lam{16}{O}
reaction~\cite{bruckner78} are shown in the center. All energies are
in MeV. The dominant
components of the wave function are shown together with the smaller
admixtures that permit proton emission to states of \lam{15}{N}}
\label{fig:lo16kpi}
\end{figure}

 At small angles in the $^{16}\mathrm{O}(\,\mathrm{K}^-,\upi^-)^{16}_{~\vlam}$O
reaction used for BNL E930, $p^{-1}p_\vlam$ $0^+$ states are strongly
excited at about 10.6 and 17.0\,MeV in excitation energy along
with a broad distribution of $s^{-1}s_\vlam$ strength centered
near 25\,MeV \cite{bruckner78}. These levels can decay by proton
emission (the threshold is at $\sim 7.8$\,MeV) to \lam{15}{N}
via $s^4p^{10}(sd)s_\vlam$ components in their wave functions.
The low-lying states of \lam{15}{N} shown in Fig.~\ref{fig:lo16kpi}
can be populated by s-wave or d-wave proton emission and higher
energy negative-parity states by p-wave emission.

 The cross section for the $0^+$ states drops rapidly with increasing
angle while the $\Delta L\! =\!1$ angular distribution rises to a maximum
near $10^\circ$~\cite{auerbach83}. The population of the excited
$1^-$ state is optimized by selecting pion angles near this maximum.
The aim of the experiment was to observe $\ugamma$-rays from the
excited $1^-$ state to both members of the ground-state doublet and
thus measure the doublet splitting. The doublet splitting is of interest
because it depends strongly on the tensor interaction. For a pure
$p_{1/2}^{-1}s_\vlam$ configuration, the combination of
parameters governing the doublet splitting is \cite{dg78}
\begin{equation}
 E(1^-_1)-E(0^-) =  -\frac{1}{3}\Delta +\frac{4}{3}S_\varLambda +8\,T \; .
\label{eq:o16gs}
\end{equation}
The measured values of the $\ugamma$-ray energies~\cite{ukai04} are
6533.9\,keV and 6560.3\,keV (with errors of $\sim 2$\,keV), giving
26.4\,keV for the splitting of the ground-state doublet. Including
recoil corrections of 1.4\,keV to the $\ugamma$-ray energies gives
6562\,keV for the excitation energy of the $1^-$ state.

 The breakdown of the contributions to the energy spacing in \lam{16}{O}
from the shell-model calculation is given in Table~\ref{tab:Lo16}
for the parameter set
\begin{equation}
\vdel= 0.430\quad \mathrm{S}_\vlam =-0.015\quad \mathrm{S}_\mathrm{N}
 = -0.350 \quad \mathrm{T}=0.0287 \; .
\label{eq:param16}
\end{equation}
These were obtained by  starting with the parameter values in
(\ref{eq:param7}) and changing T to fit the measured ground-state
doublet spacing of \lam{16}{O} and $\mathrm{S}_\mathrm{N}$ to fit
the excitation energy of the excited $1^-$ level. The most important
feature of the ground-state doublet splitting is the almost complete
cancellation between substantial contributions from T and $\vdel$
(aided by $\vlam$--$\vsig$ coupling). There is thus great sensitivity
to the value of T if $\vdel$ is fixed from other doublet spacings.

\begin{table}
\centering
\caption{Energy spacings in \lam{16}{O}. $\Delta E_\mathrm{C}$ is the
contribution of the core level spacing. The first line in each case
gives the coefficients of each of the $\vlam$N effective
interaction parameters as they enter into the spacing while the
second line gives the actual energy contributions to the spacing
in keV}
\label{tab:Lo16}
\begin{tabular}{crrrrrrr}
\hline\noalign{\smallskip}
  $J^\pi_i -J^\pi_f$ & $\Delta E_\mathrm{C}$ & $\vlam\vsig$ & $\vdel$
& S$_\vlam$ & S$_{\rm  N}$ &  T & $\Delta E$   \\
\noalign{\smallskip}\hline\noalign{\smallskip}
 $1^- - 0^-$ &  &   &  ~~$-0.380$ & 1.376  &  ~~$-0.004$ &  $7.858$  \\
 & 0 & $-30$   &  $-161$  &   $-21$ &  $1$ &   $226$  &  27  \\
$1^-_2 - 1^-_1$ &  &  & $-0.240$ & ~~$-1.252$  &  $-1.492$ &  $-0.720$ \\
 & ~~6176 & $-30$ &  $-103$  & 19 & $522$ & $-21$  &  ~~6535 \\
  $2^- - 1^-_2$ &  &   &  $0.619$ & 1.376  &  $-0.004$ &  ~~$-1.740$   \\
 & 0 & $81$   &  $266$  &   $-21$ &  $1$ &   $-50$  &  292  \\
\noalign{\smallskip}\hline
\end{tabular}
\end{table}

\begin{figure}
\centering
\includegraphics[width=10.5cm]{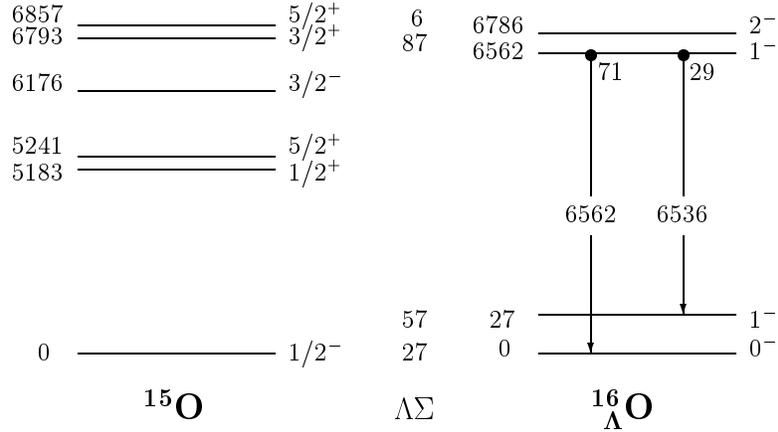}
\caption{Energy levels of \lam{16}{O} and the $^{15}$O core. The
shifts due to $\vlam$--$\vsig$ coupling are shown in the
center. All energies are in keV}
\label{fig:lo16}
\end{figure}

 Since Ref.~\cite{ukai04} was published, another peak has been found at
6758\,keV with a statistical significance of $3\sigma$. The most likely
interpretation is that it corresponds to the $2^-\to 1^-_1$ transition.
The $2^-$ level has to be excited by a weak spin-flip transition and it is
possible that states based on nearby levels of $^{15}$O, shown in
Fig.~\ref{fig:lo16}, could also be weakly excited. Accepting the first
explanation puts the $2^-$ state at 6786\,keV and implies a splitting of
224\,keV for the excited-state doublet. This is smaller than the 292\,keV
given in Table~\ref{tab:Lo16} for value of $\vdel$ used for \lamb{7}{Li}.
Reducing $\vdel$ from 0.43\,MeV to 0.33\,MeV reduces the doublet
splitting to 238 keV.  A scaling of two-body matrix elements as
$\sim \mathrm{A}^{-0.3}$ is expected for heavier nuclei but for p-shell nuclei
it is a more delicate question as could be anticipated from the discussion
of Table~\ref{tab:rms}. More evidence for a smaller value of $\vdel$ in
the latter half of the p shell comes from doublet splittings in \lam{15}{N}
and \lam{11}{B}.

\subsection{The \lam{15}{N} Hypernucleus}
\label{sec:ln15}

\begin{figure}[t]
\centering
\includegraphics[width=11.8cm]{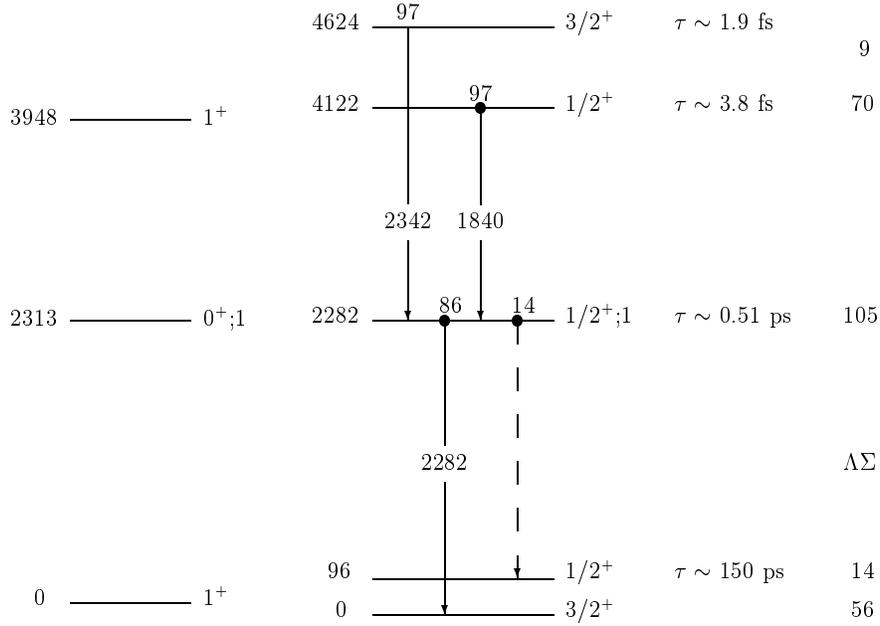}
\caption{The spectrum of \lam{15}{N} calculated from the parameters
in (\ref{eq:param15}). All energies are in keV. The levels of the $^{14}$N
core are shown on the left and the calculated lifetimes and shifts due
to $\vlam$--$\vsig$ coupling on the right}
\label{fig:ln15}
\end{figure}

 As shown in Fig.~\ref{fig:lo16kpi}, the high-energy $0^+$ states
of \lam{16}{O} populated strongly via the \Kpi reaction at forward
pion angles (and $2^+$ states at larger angles) populate states of
\lam{15}{N} by proton emission. Three $\ugamma$-ray transitions,
corresponding to the solid arrows in  Fig.~\ref{fig:ln15}
have been observed~\cite{hashtam06,ukai07}. The measured energies
are 2268, 1961, and 2442\,keV. The 2268-keV line is very sharp
without Doppler correction, indicating a long lifetime compared to
the stopping time in the target, and is identified with the
transition from the $1/2^+;1$ level to the $3/2^+$ member of the
ground-state doublet.The other two $\ugamma$-ray lines are very
Doppler broadened and therefore associated with states that have short
lifetimes.

 The excited-state doublet splitting is calculated to be 637\,keV
with the parameter set (\ref{eq:param16}). This is much larger than
the observed spacing of 481\,keV, much like the situation for the
excited-state doublet of \lam{16}{O}. The results in
Fig.~\ref{fig:ln15}, Table~\ref{tab:ln15abs}, and Table~\ref{tab:ln15}
are calculated with the parameter set
\begin{equation}
\vdel= 0.330\quad \mathrm{S}_\vlam =-0.015\quad \mathrm{S}_\mathrm{N}
 = -0.350 \quad \mathrm{T}=0.0239 \; ,
\label{eq:param15}
\end{equation}
where the value of T has been adjusted to fit the observed (26\,keV)
ground-state doublet spacing of \lam{16}{O}.

\begin{table}
\centering
\caption{Contributions of the spin-dependent $\vlam$N terms to the
binding energies of the five lowest states of \lam{15}{N} given
as the coefficients of each of the $\vlam$N effective
interaction parameters. In the $\vlam\vsig$ column, the gains
in binding energy due to $\vlam$--$\vsig$ coupling are given
in keV (same as in Fig.~\ref{fig:ln15})}
\label{tab:ln15abs}
\begin{tabular}{crrrrr}
\hline\noalign{\smallskip}
  $J^\pi_i;T$ & $\vlam\vsig$ & $\vdel$ & S$_\vlam$ & S$_\mathrm{N}$ & T \\
\noalign{\smallskip}\hline\noalign{\smallskip}
 $3/2^+_1;0$ & ~$-56$ & $-0.283$ & $~~0.780$ & 1.800 & $~~2.903$ \\
 $1/2^+_1;0$ & ~$-14$ & $~~0.457$ & $-1.457$ & 1.824 & $-6.053$ \\
 $1/2^+_1;1$ & $-105$ & $-0.022$ & $~~0.021$ & 1.816 & $-0.063$ \\
 $1/2^+_2;0$ & ~$-70$ & $-0.915$ & $-0.084$ & 0.447 & $~~0.091$ \\
 $3/2^+_2;0$ & ~~$-9$ & $~~0.452$ & $~~0.046$ & 0.481 & $-0.333$ \\
\noalign{\smallskip}\hline
\end{tabular}
\end{table}

\begin{table}
\centering
\caption{Energy spacings in \lam{15}{N}. $\Delta E_\mathrm{C}$ is the
contribution of the core level spacing. The first line in each case
gives the coefficients of each of the $\vlam$N effective
interaction parameters as they enter into the spacing while the
second line gives the actual energy contributions to the spacing
in keV. The first line of the table gives the coefficients for the
ground-state doublet in the $jj$ limit}
\label{tab:ln15}
\begin{tabular}{crrrrrrr}
\hline\noalign{\smallskip}
  $J^\pi_i -J^\pi_{f}$ & $\Delta E_\mathrm{C}$ & $\vlam\vsig$ & $\vdel$
& S$_\vlam$ & S$_\mathrm{N}$ &  T & $\Delta E$   \\
\noalign{\smallskip}\hline\noalign{\smallskip}
 $p_{1/2}^{-2}$ & & & 0.5 & $-2.0$ & 0 & $-12$ &   \\
 \phantom{;1}$1/2^+ - 3/2^+$\phantom{;1} &  &  & ~~0.740 & $-2.237$ &
0.024 &  $-8.956$ & \\
 & 0 & 42 &  244  & 33  & $-8$ & $-214$  &  96   \\
$1/2^+;1 - 3/2^+$\phantom{;1} &  &  & 0.262 & ~~$-0.752$  &  0.016
&  $-2.966$ & \\
  & 2313 & $-50$ &  86  & $11$  & $-5$ & $-71$  &  ~~2282 \\
 \phantom{;1}$3/2^+_2 - 1/2^+_2$\phantom{;1} &  &  & 1.367 & $0.130$
& 0.034 &  $-0.424$ & \\
 & 0 & 61 &  451  & $-2$  & $-12$ & $-10$  &  502  \\
\phantom{;1}$3/2^+_2 - 1/2^+;1$ &  &  & 0.474 & $0.025$ & ~~$-1.335$
& ~~$-0.271$ &\\
  & 1635 & $96$ &  156  & 0  & $467$ & $-6$  &  2342 \\
\noalign{\smallskip}\hline
\end{tabular}
\end{table}

 Table~\ref{tab:ln15abs} shows the difference between the contributions
of S$_\mathrm{N}$ for the mainly $p_{1/2}^{-2}$ and
$p_{1/2}^{-1}p_{3/2}^{-1}$ core states. In LS coupling, the $1^+$
ground state is mainly $^3$D (\ref{eq:14gs}) and the excited $1^+$ state
is mainly $^3$S. Looked at in this way, the coefficients of S$_\mathrm{N}$
for the last three states in the table arise mainly from the cross terms
between the $\mathrm{L}\!=\!0$ and $\mathrm{L}\!=\!1$ components in the core
wave functions. Small changes in the $\vlam$--$\vsig$ coupling interaction
can be used to fine tune the energy of the $1/2^+;1$ state with respect to the
$\mathrm{T}\!=\!0$ states.

 The entries for the ground-state doublet of \lam{15}{N} in
Table~\ref{tab:ln15} show a significant shift away from
the $jj$-coupling limit with the result that the higher-spin member
of the doublet is predicted to be the ground state in contrast to
the usual case for p-shell hypernuclei, including \lam{16}{O}.

 In the weak-coupling limit, the branching ratio for $\ugamma$-rays from
the $1/2^+;1$ state is 2:1 in favor of the  transition to the $3/2^+$
final state (the statistical factor from the sum over
final states). However, the transition to the $1/2^+$ state is not
observed despite the fact that the transition to the $3/2^+$ state is
very clearly observed with over 700 counts. In addition, a lifetime
estimate for the  $1/2^+;1$ level is 1.4\,ps~\cite{ukai07}, which is
very much longer than the 0.1\,ps lifetime of $0^+;1$ level in $^{14}$N.
To understand these facts requires consideration of M1 transitions
in $^{14}$N and \lam{15}{N} and this is the subject of the next
subsection.

\subsection{M1 transitions in $^{14}$N and \lam{15}{N}}
\label{sec:m1}

 The effective M1 operator can be written
\begin{equation}
\vec{\mu} = g_l^{(0)}\vec{l} + g_l^{(1)}\vec{l}\tau_3 +
 g_s^{(0)}\vec{s} +g_s^{(1)}\vec{s}\tau_3 +  g_p^{(0)}\vec{p} +
g_p^{(1)}\vec{p}\tau_3 \; ,
\label{eq:m1op}
\end{equation}
where $\vec{p} = [Y^2,\vec{s}]^1$. The values of the effective $g$ factors
that fit the M1 properties of the single-hole states in $^{15}$N and
$^{15}$O, and the states of interest in $^{14}$N are given, along
with the bare $g$ factors, in Table~\ref{tab:geff}.

\begin{table}[b]
\centering
\caption{Effective $g$ factors for M1 transitions at the end of
the p-shell. See \cite{towner87} for theoretical estimates}
\label{tab:geff}
\begin{tabular}{lrrrrrr}
\hline\noalign{\smallskip}
  & $g_l^{(0)}$ & $g_s^{(0)}$ & $g_p^{(0)}$ &
 $g_l^{(1)}$ & $g_s^{(1)}$ & $g_p^{(1)}$ \\
\noalign{\smallskip}\hline\noalign{\smallskip}
 Bare & 0.500 & ~~$0.88$ & ~~~0~ & 0.500 & ~~$4.706$ & 0 \\
 Effective & ~~0.514 & $0.76$ & 0~ & ~~0.576 & $4.120$ & ~~0.96 \\
\noalign{\smallskip}\hline
\end{tabular}
\end{table}

\begin{table}[t]
\centering
\caption{Contributions to the M1 matrix elements for $^{14}$N M1
transitions; $\mu$ is in $\mu_\mathrm{N}$ and B(M1) is in W.u.
(the M1 Weisskopf unit is $45/8\pi$ $\mu_\mathrm{N}^2$)}
\label{tab:n14m1}
\begin{tabular}{ccrrrrrr}
\hline\noalign{\smallskip}
 $J_f^\pi;T_f$ & $J_i^\pi;T_i$ & $l$ or $l\tau$ & $s$ or $s\tau$ &
 $p$ or $p\tau$ &  $\mu$/B(M1) & Exp. &  Bare g \\
\noalign{\smallskip}\hline\noalign{\smallskip}
 $1^+_1;0$ & $1^+_1;0$ & ~~0.7461 & ~$-0.3432$ & 0 & 0.403 & 0.404 &
0.328 \vspace{1.3pt}\\
  $1^+_1;0$ & $0^+;1$ & ~$-0.5070$ & ~~0.0003 & 0.2556 & 0.025
 & ~0.026(1) & ~0.077\vspace{1.3pt}\\
 $1^+_2;0$ & $0^+;1$ & $-0.5590$  & $~~3.4857$ & ~\phantom{-}$0.0304$ &
  3.50 & 3.0(9)  & 4.89\vspace{1.3pt}\\
$1^+;0$ & $2^+;1$ & $~~0.2282$  & $-4.1491$ & $0.1653$ &
  1.13 & 0.99  & 1.65\vspace{1.3pt}\\
$2^+;0$ & $2^+;1$ & $~~0.1651$  & $~~3.7665$ & $0.1884$ &
  2.26 & 2.29  & 2.64 \\
\noalign{\smallskip}\hline
\end{tabular}
\end{table}

  The B(M1) value is given by
\begin{equation}
\mathrm{B(M1)} = \frac{3}{4\pi}\ \frac{2J_f+1}{2J_i+1}\ \mathrm{M}^2 \: ,
\label{eq:bm1}
\end{equation}
where
\begin{equation}
 \mathrm{M} = \langle f||\vec{\mu}^{(0)}||i\rangle + \langle
T_i\, M_T\,1\,0\,|T_f\,M_T\rangle\langle f||\vec{\mu}^{(1)}||i\rangle \: .
\label{eq:mm1}
\end{equation}

 Contributions to the M1 matrix elements for $^{14}$N M1
transitions are given in Table~\ref{tab:n14m1} for an interaction
fitted in the manner described following (\ref{eq:14gs}) and
(\ref{eq:14GT}). The important thing to notice, apart from the fact that
the effective operator with this set of wave functions does describe
the data well, is that the $0^+;1\to gs$ transition is hindered
while the $1^+_2;0\to 0^+;1$ transition is strong. In the former case,
the $<\sigma\tau>$ matrix element is $\sim 0$ by construction
(\ref{eq:14GT}) while in the latter it is very strong reflecting
the allowed $^3\mathrm{S}\to {^1\mathrm{S}}$ nature of the transition. 
Also, the sign of the two matrix elements is different.

 To see what this means for the M1 transitions de-exciting the
$1/2^+;1$ state of \lam{15}{N}, the most important components
of the shell-model wave functions for \lam{15}{N} are listed in
Table~\ref{tab:ln15wf}. The small $1^+_2;0\!\times\! s_\varLambda$
admixtures in the wave functions for the ground-state doublet
will clearly lead to cancellations in the relevant M1 matrix elements
because they bring in a large positive matrix element while the
M1 matrix element between the large components is small and negative.

 The general expression for electromagnetic matrix elements between
hypernuclear basis states is
\begin{eqnarray}
 \lefteqn{\langle\left( J_\mathrm{c} T_\mathrm{c} s_\mathrm{Y}t_\mathrm{Y}\right)J_fT_f
||M||\left( J_\mathrm{c}' T_\mathrm{c}' s_\mathrm{Y}'t_\mathrm{Y}'\right)J_iT_i\rangle }
 \nonumber \\
& & = \delta_\mathrm{YY'}
\left(\begin{array}{lll} J_\mathrm{c}' & \Delta J & J_\mathrm{c} \\
1/2 & 0 & 1/2 \\
J_i & \Delta J  & J_f \end{array} \right)
\left(\begin{array}{lll} T_\mathrm{c}' & \Delta T & T_\mathrm{c} \\
t_\mathrm{Y} & 0 & t_\mathrm{Y} \\
T_i & \Delta T  & T_f \end{array} \right)
\langle J_\mathrm{c} T_\mathrm{c} ||M^{\Delta J\Delta T}_\mathrm{c} ||
J_\mathrm{c}' T_\mathrm{c}'\rangle \nonumber \\
& &  ~+ \delta_\mathrm{cc'}
\left(\begin{array}{lll} J_\mathrm{c} & 0 & J_\mathrm{c} \\
1/2 & \Delta J  & 1/2 \\
J_i & \Delta J  & J_f \end{array} \right)
\left(\begin{array}{lll} T_\mathrm{c} & 0 & T_\mathrm{c} \\
t_\mathrm{Y}' & \Delta T  & t_\mathrm{Y} \\
T_i & \Delta T  & T_f \end{array} \right)
\langle s_\mathrm{Y}t_\mathrm{Y}||M^{\Delta J\Delta T}_\mathrm{Y} ||
s_\mathrm{Y}'t_\mathrm{Y}' \rangle \; .
\label{eq:hypm1}
\end{eqnarray}

\begin{table}[t]
\centering \caption{Excitation energies and weak-coupling wave
functions for \lam{15}{N}} \label{tab:ln15wf}
\begin{tabular}{ccl}
\hline\noalign{\smallskip}
 $J^\pi_n;T$ & $E_x$ (keV) & ~~~~~~~~~~~~~~~~~Wave function \\
\noalign{\smallskip}\hline\noalign{\smallskip}
 $3/2^+_1;0$ &  ~~~0 & $~~0.9985\,1^+_1;0\times s_\varLambda\ +\
 0.0318\,1^+_2;0\times s_\varLambda\ +\ 0.0378\,2^+_1;0\times s_\varLambda$
\vspace{1.3pt}\\
 $1/2^+_1;0$ &  ~~96 & $~~0.9986\,1^+_1;0\times s_\varLambda\ +\
 0.0503\,1^+_2;0\times s_\varLambda$ \vspace{1.3pt}\\
 $1/2^+_1;1$ &  2282 & $~~0.9990\,0^+_1;1\times s_\varLambda\ +\
 0.0231\,1^+_1;1\times s_\varLambda\ +\ 0.0206\,0^+_2;1\times s_\varLambda$
\vspace{1.3pt}\\
  & & $-0.0261\,0^+_1;1\times s_\varSigma$ \vspace{1.3pt}\\
 $1/2^+_2;0$ &  4122 & $-0.0502\,1^+_1;0\times s_\varLambda\ +\
 0.9984\,1^+_2;0\times s_\varLambda$ \vspace{1.3pt}\\
 $3/2^+_2;0$ &  4624 & $-0.0333\,1^+_1;0\times s_\varLambda\ +\
 0.9984\,1^+_2;0\times s_\varLambda\ +\ 0.0363\,2^+_1;0\times s_\varLambda$ \\
\noalign{\smallskip}\hline
\end{tabular}
\end{table}

The two important $g$ factors in the hyperonic sector are
$g_\vlam \!= -1.226$\,$\mu_\mathrm{N}$ and
$g_{\vlam\vsig} \!= \!3.22$\,$\mu_\mathrm{N}$
(the $g$ factors for the $\vsig$ hyperons are included in the calculations).
For hypernuclear doublet transitions in the weak-coupling limit,
\begin{eqnarray}
\vec{\mu} & = & g_\mathrm{c}\vec{J}_\mathrm{c} + g_\vlam\vec{J}_\vlam
\nonumber \\
 & = & g_\mathrm{c}\vec{J}  + (g_\vlam - g_\mathrm{c})\vec{J}_\vlam
\label{eq:ghypdoub}
\end{eqnarray}
can be used to obtain a simple expression for the matrix element
in terms of $g_\vlam - g_\mathrm{c}$ as an overall multiplicative
factor~\cite{dg78}.

 The important contributions for M1 decays from the $1/2^+;1$
state in \lam{15}{N} are shown in Table~\ref{tab:ln15m1}. The
strong cancellation resulting from the small $1^+_2\!\times\! s_\vlam$
admixtures is evident. Even the small $\vsig$ admixtures contribute
to the cancellation. The cancellation is stronger for the transition
to the $1/2^+$ member of the ground-state doublet. The reason for this
can be seen from Table~\ref{tab:offdiag}. Namely, the largest contributions
to the off-diagonal matrix elements come from S$_\mathrm{N}$ and T and
add for the $1/2^+$ state and cancel for the $3/2^+$ state.
\begin{table}
\centering
\caption{Important contributions for M1 decays from the $1/2^+;1$
state in \lam{15}{N}}
\label{tab:ln15m1}
\begin{tabular}{clrr}
\hline\noalign{\smallskip}
 $1/2^+;1\to 3/2^+;0$ & ~~large component & ~~$0.9979\times 0.9988\times
(-0.251)$  &  $-0.250$ \\
 & ~~$1^+_2\times s_\vlam$ admixture & ~~$0.0318\times 0.9988\times
(\phantom{-}2.957)$  &  ~~$+0.095$ \\
 & ~~$\varSigma$ admixture &            & $+0.011$ \\
 & ~~Partial sum & & $-0.137$ \\
\noalign{\smallskip}\hline\noalign{\smallskip}
$1/2^+;1\to 1/2^+;0$ &  ~~large component & $0.9983\times 0.9988\times
(-0.251)$   &  $-0.250$ \\
 & ~~$1^+_2\times s_\vlam$ admixture & $0.0545\times 0.9988\times
(\phantom{-}2.957)$  &  $+0.161$ \\
 & ~~$\varSigma$ admixture &            & $+0.008$ \\
 & ~~Partial sum  & & $-0.081$ \\
\noalign{\smallskip}\hline
\end{tabular}
\end{table}
\begin{table}
\centering
\caption{Coefficients of the $\vlam$N interaction parameters
in the off-diagonal matrix elements between the $1^+_1;0\times s_\vlam$
and $1^+_2;0\times s_\varLambda$ basis states in \lam{15}{N}
and the $1^-$ states in \lam{16}{O}. The second line gives the energy
contributions in MeV}
\label{tab:offdiag}
\begin{tabular}{crrrrr}
\hline\noalign{\smallskip}
  $J^\pi$ & $\vdel$ & S$_\vlam$ & S$_\mathrm{N}$ & T & ME  \\
\noalign{\smallskip}\hline\noalign{\smallskip}
  $1/2^+$ & 0.1275 & ~~$-0.1275$ & 0.4851 & ~~$-4.0664$ & \\
 & $~~0.0421$ & $~~0.0019$ & ~~$-0.1698$ & $-0.0972$ & ~~$-0.223$ \\
  $3/2^+$ & ~~$-0.0637$ & 0.0637 & 0.4851 & 2.0332 & \\
 & $-0.0210$ & $-0.0010$ & $-0.1698$ & $~~0.0486$ & $-0.143$ \\
  $1^-$ & 0.4714 & -0.4714 & 0. & 1.4142 & \\
 & $0.1556$ & $0.0071$ &  0.  & $~~0.0338$ & \phantom{$-$}$0.196$ \\
\noalign{\smallskip}\hline
\end{tabular}
\end{table}

Finally, the M1 transition data for \lam{16}{O} and \lam{15}{N}
are collected in Table~\ref{tab:1516m1}, mainly to emphasize the
weakness of the M1 transitions from the $1/2^+;1$ level of \lam{15}{N}.

\begin{table}
\centering
\caption{M1 transition strengths in \lam{16}{O} and \lam{15}{N}}
\label{tab:1516m1}
\begin{tabular}{ccrrrr}
\hline\noalign{\smallskip}
 \ $J_f^\pi;T_f$ & ~~$J_i^\pi;T_i$ & $E_\ugamma$ (keV) & ~B(M1) (W.u.) &
~$\ugamma$ branch (\%) & ~~lifetime \\
\noalign{\smallskip}\hline\noalign{\smallskip}
 $0^-_1;1/2$ & ~~$1^-_2;1/2$ & 6562 & 0.336 & 72.5 & 0.24 fs \\
 $1^-_1;1/2$ & ~~$1^-_2;1/2$ & 6535 & 0.129 & 27.5 &  \\
 $0^-_1;1/2$ & ~~$1^-_1;1/2$ & ~~26 & 0.176 & weak & 10 ns \\
 $3/2^+_1;0$ & ~~$1/2^+;1$ & 2268 & $4.55\times 10^{-3}$ & 86 & 0.51 ps \\
 $1/2^+_1;0$ & ~~$1/2^+;1$ & 2172 & $8.89\times 10^{-4}$ & 14 &  \\
 $3/2^+_1;0$ & ~~$1/2^+_1;0$ & ~~96 & 0.240 & weak/$\ugamma$ &  150 ps \\
 $1/2^+;1$ & ~~$3/2^+_2;0$ & 2442 & 1.133 & 96.9 & 1.9 fs \\
 $1/2^+;1$ & ~~$1/2^+_2;0$ & 1961 & 1.080 & 97.4 & 3.8 fs \\
\noalign{\smallskip}\hline
\end{tabular}
\end{table}

\subsection{The \lam{10}{B}, \lam{12}{C}, and \lam{13}{C} Hypernuclei}
\label{sec:lc12}

 It was noted in Sect.~\ref{sec:911} that $^9$Be/$^9$B and $^{11}$B/$^{11}$C
have similar structure, as is evident from Fig.~\ref{fig:911}. The
hypernuclei \lam{10}{B} and \lam{12}{C} will have $2^-$/$1^-$ ground-state
doublets with $1^-$ as the ground state (this is known experimentally
for \lam{12}{B}~\cite{davis86}). However, there are considerable
differences in how these levels can be studied experimentally.
In \lam{10}B, only the states of the ground-state doublet are particle-stable
because, as Fig.~\ref{fig:lbe9p} shows, the neutron threshold is at
2.00\,MeV while the proton threshold is at 9.26\,MeV in \lam{12}{C}.
Fig.~\ref{fig:lbe9p} also shows that the $2^-$ state of \lam{10}{B}
is populated by non-spin-flip transitions from the $3^+$ ground state
of $^{10}$B. The resulting $\ugamma$-ray transition was first searched for
in~\cite{chrien90} without success, an upper limit of 100\,keV being put
on the doublet spacing (in BNL E930, the transition was also looked for and
not found at roughly the same limit). In \lam{12}{C}, it is the
$1^-$ ground state that is populated by non-spin-flip transitions from
a $^{12}$C target and the doublet spacing is best investigated by
looking for transitions from higher bound states of \lam{12}{C}. This
approach was tried in KEK E566 and the data is still under analysis.

 The similarity of the contributions from the spin-dependent $\vlam$N
interaction to the two ground-state doublets is shown in
Table~\ref{tab:c12b10a} for a calculation using the parameters of
(\ref{eq:param7}) and the fitted p-shell interaction used for
Fig~\ref{fig:hsu4}. If the parameters of  (\ref{eq:param15}) are
used the ground-state doublet spacings for \lam{10}{B} and \lam{12}{C}
drop to 121\,keV and 150\,keV, respectively.

\begin{table}[t]
\centering
\caption{Coefficients of the $\vlam$N interaction parameters
for the $2^-$/$1^-$ ground-state doublet separations of
\lam{10}{B} and \lam{12}{C}. The energy contributions from
$\vlam$--$\vsig$ coupling and the doublet splitting $\Delta E$ are in keV}
\label{tab:c12b10a}
\begin{tabular}{ccccccc}
\hline\noalign{\smallskip}
   & $\vlam\vsig$ & $\vdel$ & S$_\vlam$ & S$_\mathrm{N}$ & T & $\Delta E$ \\
\noalign{\smallskip}\hline\noalign{\smallskip}
\lam{12}{C} & ~$\phantom{-}58$ & ~~0.540 & ~~1.44 & ~~0.046 & ~$-1.72$
& ~~191 \\
\lam{10}{B} & ~$-15$ &  ~~0.578 & ~~1.41  &  ~~0.013 & ~$-1.07$ & ~~171\\
\noalign{\smallskip}\hline
\end{tabular}
\end{table}

 The most notable point to be taken from Table~\ref{tab:c12b10a} is
that the $\vlam$--$\vsig$ coupling increases the doublet spacing
in \lam{12}{C} and reduces it in \lam{10}{B}. The reason for this
is that spin-spin matrix element for the $\vlam$N interaction
depends on an isoscalar one-body density-matrix element of the nuclear
spin operator for the core while the corresponding matrix element
for  $\vlam$--$\vsig$ coupling depends on an isovector one-body
density-matrix element of the nuclear spin operator for the core [see
(\ref{eq:manybody})]. The isoscalar and isovector matrix elements
are both large but they have opposite relative sign for the two
hypernuclei (this is a type of particle-hole symmetry for the
$\mathrm{K}\!=\!3/2$ Nilsson orbit). The coupling matrix elements are broken down
in Table~\ref{tab:c12b10b}. The ``diagonal'' matrix elements
involving the $3/2^-$ core states contain a contribution of 1.45\,MeV
from $\overline\mathrm{V}\,'$ (\ref{eq:lsparam7}) and the contribution from
$\vdel'$ produces the shifts from this value. If it were not from
the contribution to the energy shifts from the $1/2^-\times\vsig$
configuration (the $1/2^-$ and $3/2^-$ core states both have $\mathrm{L}\!=\!1$),
there would be a much larger effect on the relative ground-state
doublet spacings in \lam{10}{B} and \lam{12}{C}.

\begin{table}
\centering
\caption{Matrix elements (in MeV) coupling $\vsig$ configurations with
the members of the $3/2^-\times\vlam$ ground-state doublets in
\lam{10}{B} and \lam{12}{C}. The energy shifts caused by these
couplings are given in keV}
\label{tab:c12b10b}
\begin{tabular}{ccccc}
\hline\noalign{\smallskip}
 & $J^\pi$ & ~~$3/2^-\times\vsig$~~ & ~$1/2^-\times\vsig$~ &
~~$\vlam\vsig$ shift \\
\noalign{\smallskip}\hline\noalign{\smallskip}
\lam{10}{B} &  $1^-$ & 0.55  &  $\phantom{-}1.47$  & 34\\
            &   $2^-$ & 1.95 &   & 49 \\
\lam{12}{C} &  $1^-$ & 1.92   & $-1.35$  & 98  \\
                      &   $2^-$ & 1.13  &  & 40   \\
\noalign{\smallskip}\hline
\end{tabular}
\end{table}

\begin{figure}
\centering
\includegraphics[width=9.5cm]{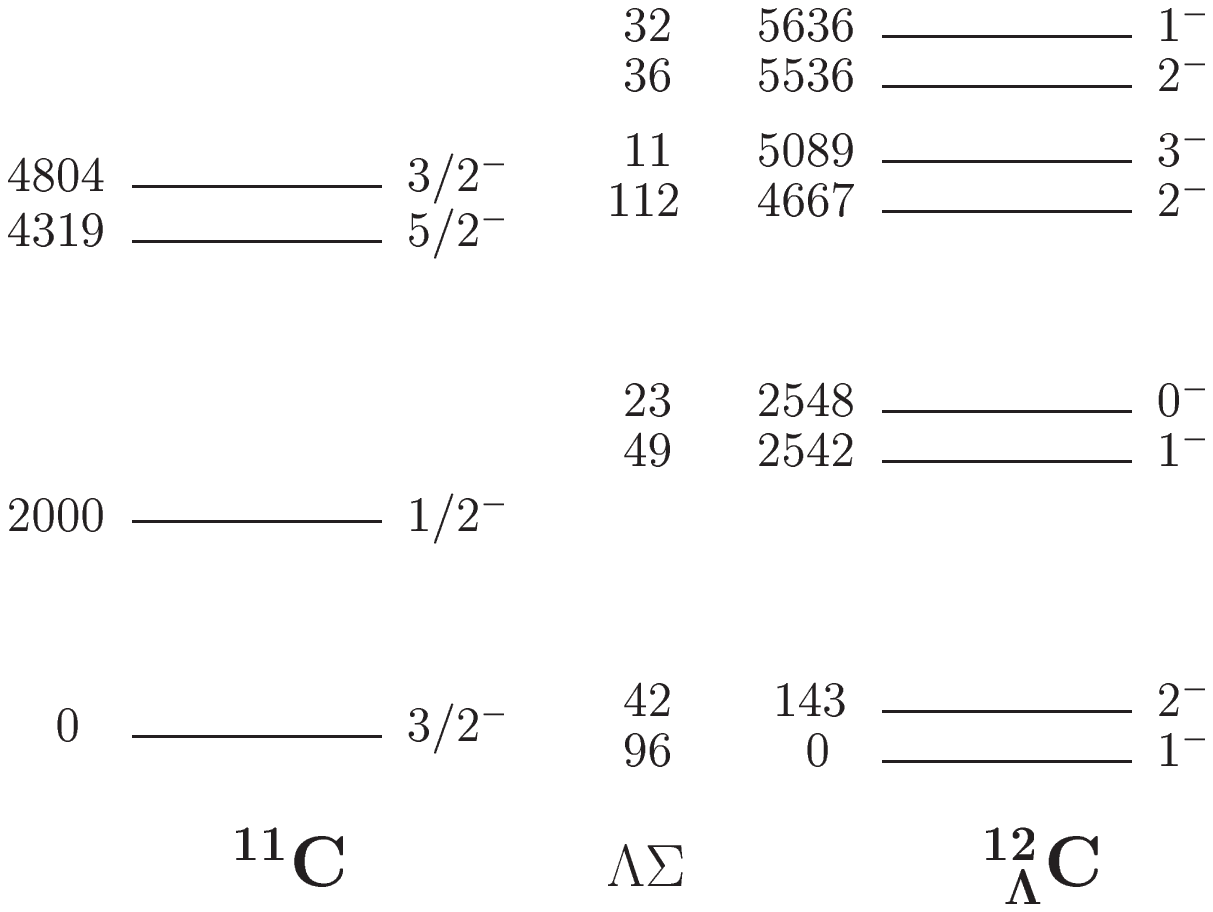}
\caption{The spectrum of \lam{12}{C} calculated from the parameters
in (\ref{eq:param7}). The levels of the $^{11}$C core are shown on the
left and the calculated shifts due to $\vlam$--$\vsig$ in the center.
All energies are in keV}
\label{fig:lc12}
\end{figure}

 Apart from the effect of $\vlam$--$\vsig$
coupling, several of the coefficients in Table~\ref{tab:c12b10a} are
sensitive to the model of the p-shell core. For example, the
ground states of the core nuclei tend to be characterized by a good K
value and this involves a mixing of L values as noted in, and following,
(\ref{eq:c11gs}). For $\mathrm{L}\!=\!1$, the coefficient of $\vdel$ 
contributing to the doublet spacing is 2/3 whereas for 
$\mathrm{L}\!=\!2$ the coefficient
is $-2/5$. For the wave function in (\ref{eq:c11gs}), the coefficient
is the $6/13 \sim 0.46$. The CK816 interaction gives a coefficient
close to this value and the results of a calculation for \lam{12}{C}
with this interaction and the parameter set (\ref{eq:param7}) are
shown in Fig.~\ref{fig:lc12} and Table~\ref{tab:lc12}.

\begin{table}
\centering
\caption{Energy spacings in \lam{12}{C}. $\Delta E_\mathrm{C}$ is the
contribution of the core level spacing. The first line in each case
gives the coefficients of each of the $\vlam$N effective
interaction parameters as they enter into the spacing while the
second line gives the actual energy contributions to the spacing
in keV}
\label{tab:lc12}
\begin{tabular}{crrrrrrr}
\hline\noalign{\smallskip}
  $J^\pi_i -J^\pi_f$ & $\Delta E_\mathrm{C}$ & $\vlam\vsig$ & $\vdel$
& S$_\vlam$ & S$_\mathrm{N}$ &  T & $\Delta E$   \\
\noalign{\smallskip}\hline\noalign{\smallskip}
 $2^-_1 - 1^-_1$ & & & ~~0.463 & 1.518 & 0.030 & ~$-2.078$ & \\
  & 0 & ~~54 & 199 & $-23$ & $-12$ & $-62$ & 143   \\
 $1^-_2 - 1^-_1$ & & & 0.315 & 1.150 & ~$-1.104$ & 0.635 & \\
    & ~~2000 & 45 & 136  & $-17$  & 430 & 19 & ~~2548  \\
  $1^-_3 - 1^-_1$ & & & 0.372 & ~$-0.385$ & $-1.647$ & 0.561 & \\
   & 4804 & 64 & 160  & 6  & 642 & 17 & 5536   \\
\noalign{\smallskip}\hline
\end{tabular}
\end{table}

 The non-observation of the ground-state doublet spacing in \lam{10}{B}
is an important problem. A number of p-shell interactions give a
smaller coefficient for $\vdel$ (due to more L mixing) or larger
coefficient of T, both of which lead to a reduction in the doublet
spacing. A better understanding of how the parameters vary with
mass number and, indeed, whether the parametrization in use is
sufficient are also important questions.

 There have been many experiments using a $^{12}$C target~\cite{hashtam06}
and many show excitation strength in the region of the excited
$1^-$ states. For example, in the \piK reaction with a thin carbon
target, the second $1^-$ state is found at 2.5\,MeV~\cite{hotchi01}.
Table~\ref{tab:lc12} shows that the S$_\mathrm{N}$ parameter is
mainly responsible for raising the excitation energy above the core
spacing of 2\,MeV. Recently, excited states of \lam{12}{B} have been
observed with better resolution via the
$(\mathrm{e},\mathrm{e'}\mathrm{K}^+\!)$ reaction~\cite{iodice07}.

 A similar effect of S$_\mathrm{N}$ is seen in Table~\ref{tab:lc13}
for \lam{13}{C} where the excited $3/2^+$ state built on the 4.44-MeV
$2^+$ state of $^{12}$C
(cf. Fig.~\ref{fig:lbe9}) is seen at 4.880(20)\,MeV in a $\ugamma$-ray
experiment using NaI detectors~\cite{kohri02} and at 4.85(7)\,MeV in
KEK E336 via the \piK reaction~\cite{hashtam06}. Note that, as for
\lamb{9}{Be}, the effects of $\vlam$--$\vsig$ coupling are small.
In contrast to \lamb{9}{Be}, the coefficients of  S$_\mathrm{N}$
are large. This is because the $^{12}$C core states, while
still having dominantly [44] spatial symmetry, have substantial
[431] components with $\mathrm{S}\!=\!1$ (a low 68\% [44] and 25\% [431] in the
ground state for the WBP interaction~\cite{warburton92} used
for Table~\ref{tab:lc13} but typically $\sim 79$\% [44] for the
Cohen and Kurath interactions).

 The 4.88-MeV $\ugamma$-ray was actually a by-product of an 
experiment~\cite{kohri02} designed to measure the spacing of 
$1/2^-$ and $3/2^-$ states at $\sim 11$\,MeV in \lam{13}{C}
(B$_\vlam\!=\!11.67$\,MeV is the lowest particle threshold).
To a first approximation, the two states are pure $p_\vlam$ 
single-particle states. In this case, and with harmonic oscillator
wave functions, the spacing produced by the interaction of the
$p_\vlam$ interacting with the filled s shell is related to 
S$_\vlam$ (with a coefficient of $-6$) because both depend on 
the same Talmi integral $I_1$~\cite{mgdd85}. 
However, as noted above, the $^{12}$C 
core is not by any means pure $\mathrm{L}\!=\!0$, $\mathrm{S}\!=\!0$
which means that components of the $\vlam$N effective interaction other
than the $\vlam$ spin-orbit interaction, particularly the tensor
interaction, play important roles in the small spacing of 
152\,keV~\cite{millener01}. In addition, the loose binding of the
$p_\vlam$ orbit is important (the harmonic oscillator approximation
is not good), as is configuration mixing produced by the 
quadrupole-quadrupole component in the $p_\mathrm{N}p_\vlam$
interaction~\cite{millener01,auerbach83}.

\begin{table}[t]
\centering \caption{Energy spacings in \lam{13}{C}. Coefficients
of the $\vlam$N effective interaction parameters for the $1/2^+$
ground state and $3/2^+$ excited state are given followed by the
difference and the actual energy contributions in keV}
\label{tab:lc13}
\begin{tabular}{crrrrrrr}
\hline\noalign{\smallskip}
  $J^\pi$ & $\Delta E_\mathrm{C}$ & $\vlam\vsig$ & $\vdel$
& S$_\vlam$ & S$_\mathrm{N}$ &  T & $\Delta E$   \\
\noalign{\smallskip}\hline\noalign{\smallskip}
 $1/2^+$ &     & ~~27 & ~~$-0.016$ & $0.016$ & $2.421$ & ~~$-0.049$ & \\
 $3/2^+$ &     & ~~18 & $-0.045$ & ~~$-1.455$ & $1.430$ & $-0.929$ & \\
$3/2^+-1/2^+$ & ~~4439 & & ~$-0.029$ & ~~$-1.471$ & $-0.991$ &
$-0.880$ & \\
  & & 9 & $-12$  &  22  &  $386$  &   $-28$  &  ~~4803   \\
\noalign{\smallskip}\hline
\end{tabular}
\end{table}

\subsection{The \lam{11}{B} Hypernucleus}
\label{sec:lb11}

 The \lam{11}{B} hypernucleus has a rather complex spectrum because
the $^{10}$B core has many low-lying p-shell levels, as shown in
Fig.~\ref{fig:b10be10}. The $\ugamma$-decay properties of these levels
have been very well studied~\cite{tunl}. Furthermore, the lowest
particle threshold (proton) in \lam{11}{B} is at 7.72\,MeV which
means that the hypernuclear states based on the p-shell states
of $^{10}$B up to 6\,MeV or so are expected to be particle stable
and thus could be seen via their $\ugamma$ decay if they could be
populated strongly enough.

\begin{figure}
\centering
\includegraphics[width=10.5cm]{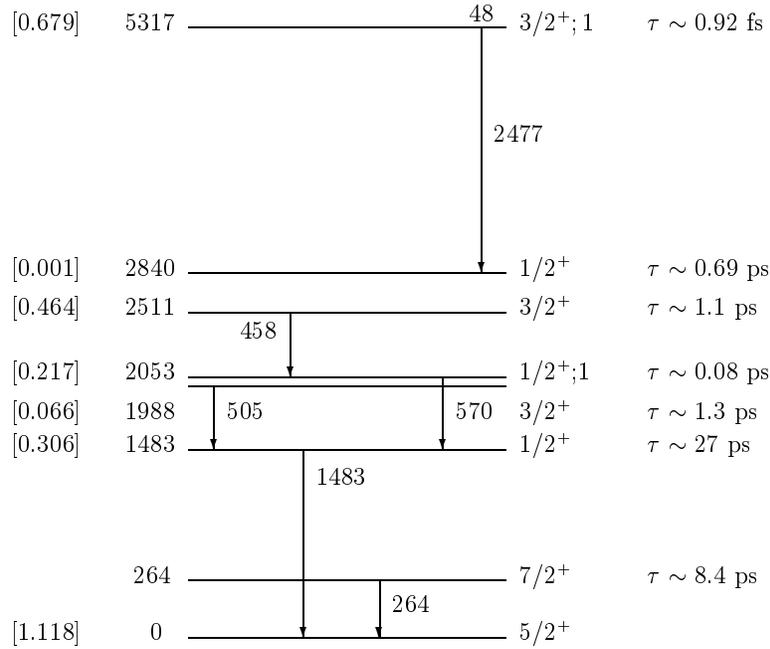}
\caption{The spectrum of \lam{11}{B} based on the six observed
$\ugamma$-ray transitions. All energies are in keV. The placements of the
264-keV, 1482-keV, and 2477-keV transitions are  well founded. The
placements of the other three $\ugamma$-rays are more speculative.
The formation factors for the \piK reaction on the left and the
lifetimes on the right are from the shell-model calculation}
\label{fig:lb11}
\end{figure}

 A shell-model calculation for \lam{11}{B} was made using the p-shell
interaction of Barker~\cite{barker81} who made some changes to one of
the Cohen and Kurath interactions~\cite{ck65} to improve the
description of electromagnetic transitions in $^{10}$B. The
strengths for formation via non-spin-flip transitions and the
electromagnetic matrix elements for decay were calculated for all
the bound p-shell hypernuclear states of \lam{11}{B} (i.e, up to
the states based on the 5.92-MeV $2^+;0$ level of $^{10}$B).
The $\ugamma$-ray cascade was followed from the highest levels,
summing the direct formation strength and the feeding by $\ugamma$
rays from above. The conclusion was that perhaps as many as eight
transitions would contain enough intensity to be seen in an
experiment with the Hyperball. The formation strengths on the
left side of Fig.~\ref{fig:lb11} show that the most strongly
formed excited state is expected to be the $3/2^+$ level based on
the 5.16-MeV $2^+;1$ state of $^{10}$B, followed by a number of states
based on the low-lying $1^+;0$ and $0^+;1$ states. The lowest
$1/2^+;0$ level, originally predicted at 1.02\,MeV, acts as a collection
point for the $\ugamma$-ray cascade. The predicted $\ugamma$ width at this
energy corresponds to a lifetime of $\sim 250$\,ps (the $1^+$ state of
$^{10}$B has to decay by an E2 transition and has a lifetime of 1\,ns)
which is comparable with the expected lifetime for weak decay.

 In the subsequent experiment, KEK E518, six $\ugamma$ rays were
seen~\cite{miura05,tamura05,hashtam06}. Figure \ref{fig:lb11}
shows an attempt to construct a level scheme for \lam{11}{B}
from a combination of the experimental information and the results
of the shell-model calculation. The theoretical energies and the
contributions from YN effective interactions are given in
Table~\ref{tab:lb11} for the parameter set (\ref{eq:param15}).
Apart from the $\ugamma$-ray energies, the experimental information
includes relative intensities and some estimate of the lifetimes
from the degree of Doppler broadening.

\begin{table}[t]
\centering
\caption{Contributions of the spin-dependent $\vlam$N terms to the
binding energies of the eight levels of \lam{11}{B} shown in
Fig.~\ref{fig:lb11} given as the coefficients of each of the
$\vlam$N effective interaction parameters. The theoretical
excitation energies and the gains in binding energy due to
$\vlam$--$\vsig$ coupling are given in keV}
\label{tab:lb11}
\begin{tabular}{crrrrrr}
\hline\noalign{\smallskip}
  $J^\pi;T$ & E$_x$ &$\vlam\vsig$ & $\vdel$ & S$_\vdel$ & S$_\mathrm{N}$ & T \\
\noalign{\smallskip}\hline\noalign{\smallskip}
 $5/2^+;0$  &  ~~~0 & ~66 &  $-0.616$ & ~$-1.377$  &  $~~1.863$ &  $~~1.847$ \\
 $7/2^+;0$  &  ~266 & ~11 &  $~~0.409$ & $~~1.090$  & $~~1.890$ &  ~$-1.512$ \\
 $1/2^+;0$  &  ~968 & ~71 &  ~$-0.883$ & $-0.116$  &  $~~0.746$ &  $~~0.243$ \\
  $3/2^+;0$  &  1442 & ~12 &  $~~0.403$ & $~~0.094$  & $~~0.872$ &  $-0.194$ \\
  $1/2^+;1$  &  1970 & ~93 &  $-0.007$ & $~~0.008$  &  $~~1.543$ &  $-0.013$ \\
  $3/2^+;0$  &  2241 & ~46 &  $-0.266$ & $~~0.754$  &  $~~1.536$ &  $-1.264$ \\
  $1/2^+;0$  &  2554 & ~35 &  $~~0.333$ & $-1.333$  &  $~~1.674$ & $~~2.639$ \\
 $3/2^+;1$  & ~~5366 & ~~103 & $-0.203$ & $-1.293$  & $~~1.519$ &  $~~0.598$ \\
\noalign{\smallskip}\hline
\end{tabular}
\end{table}

  The strongest $\ugamma$ ray in the spectrum was found at 1483\,keV
and it is very sharp implying a long lifetime. Despite the unexpectedly
high energy, it is natural to associate this $\ugamma$ ray with the
de-excitation of the lowest $1/2^+;0$ level. The 2477-keV $\ugamma$
ray shows up after the Doppler-shift correction and it too has a natural
assignment in Fig.~\ref{fig:lb11}. It is a 1.1\,W.u. isovector M1 transition
between states with $\mathrm{L}\!=\!2$ and K$_\mathrm{L}\!=\!0$. The 264-keV
line is now known
to be due to the ground-state doublet transition (0.2\,W.u.), having been seen
following proton emission from \lam{12}{C}~\cite{tamura} (this is the reason
for showing a calculation using the parameter set (\ref{eq:param15}) with
$\vdel\!=\!0.33$\,MeV). The placement of the other three $\ugamma$ transitions
in Fig.~\ref{fig:lb11} is speculative, although the intensities and
lifetimes match the theoretical estimates quite well. The $1/2^+;1\to
1/2^+;0$ transition is 2.0\,W.u. M1 transition between states with
$\mathrm{L}\!=\!0$.

 The most glaring
discrepancy is that the shell-model calculation greatly underestimates
the excitation energies of the two doublets based on the $1^+;0$
levels of $^{10}$B. From Table~\ref{tab:lb11}, it can be seen that
S$_\mathrm{N}$ does raise the energies of these doublets with respect
to the ground-state doublet but not nearly enough. The shell-model
calculation is in fact quite volatile with respect to the p-shell
wave functions for the $1^+;0$ core levels. There is also mixing
of the members of these two doublets and this is evident from the
difference between the coefficients of  S$_\mathrm{N}$ for the
doublet members.

\section{Summary and Outlook}
\label{sec:summary}

  The era of Hyperball experiments at KEK and BNL between 1998 and 2005
has provided accurate energies for about 20 $\ugamma$-ray transitions
in p-shell hypernuclei, the number in each hypernucleus being
five for \lamb{7}{Li}, two for \lamb{9}{Be}, six for \lam{11}{B}, three
for \lam{15}{N}, and three for \lam{16}{O}. Data from the last
experiment, KEK 566, on a $^{12}$C target using the upgraded Hyperball-2
detector array is still under analysis but there is evidence for one
$\ugamma$-ray transition in \lam{12}{C} and two $\ugamma$ rays from
\lam{11}{B} have been seen following proton emission from the region of
the $p_\vlam$ states of \lam{12}{C}~\cite{tamura}. Several electromagnetic
lifetimes have been measured by the Doppler shift attenuation method or
lineshape analysis, and many estimates of, or limits on, lifetimes have been
made based on the Doppler broadening of observed $\ugamma$ rays. In addition,
two measurements of $\ugamma\ugamma$ coincidences have been made, for
the $7/2^+\to5/2^+\to 1/2^+$ cascade in \lamb{7}{Li} (471-keV and 2050-keV
$\ugamma$ rays) and the $3/2^+\to1/2^+\to 3/2^+$ cascade in \lam{15}{N}
(2442-keV and 2268-keV $\ugamma$ rays).

  With the exception of transitions in \lam{11}{B} that most likely involve
levels based on the two lowest $1^+$ states of $^{10}$B, the $\ugamma$-ray
data can be accounted for by shell-model calculations that include
both $\vlam$ and $\vsig$ configurations with p-shell cores. The
spin-dependence of the effective $\vlam$N interaction appears to be
well determined. The singlet central interaction is more attractive
than the triplet as evidenced by the value $\vdel\! =\! 0.43$ MeV needed
to fit the 692-keV ground-state doublet separation in \lamb{7}{Li}
(and the 471-keV excited-state doublet spacing). In \lamb{7}{Li},
the contribution from $\vlam$--$\vsig$ coupling is $\sim 12$\% of the
contribution from the $\vlam$N spin-spin interaction in contrast to
the $0^+$, $1^+$ spacings in the $\mathrm{A}\! =\!4$ hypernuclei, where
the contributions are comparable
in magnitude. The $\vlam$N interaction parameters do exhibit a dependence
on nuclear size. For example, the spacings of the excited-state doublets
in \lam{16}{O} ($1^-$, $2^-$) and \lam{15}{N} ($1/2^+$, $3/2^+$), and
the ground-state doublet in \lam{11}{B} ($5/2^+$, $7/2^+$) require
$\vdel\sim 0.32$\,MeV. Given $\vdel$, the tensor interaction strength
T is well determined ($\sim 0.025$\,MeV) by the ground-state
doublet ($0^-$, $1^-$) spacing in \lam{16}{O} because of the sensitivity
provided by a strong cancellation involving T and $\vdel$. The
$\vlam$-spin-dependent spin-orbit strength S$_\vlam$ is constrained to be
very small ($\sim -0.015$\, MeV) by the excited-state doublet
spacing in \lamb{9}{Be} ($3/2^+$, $5/2^+$) . Finally, substantial effects
of the nuclear-spin-dependent spin-orbit parameter
S$_\mathrm{N}\sim -0.4$\,MeV, which effectively augments the nuclear
spin-orbit interaction in changing the spacing of core levels in
hypernuclei, are seen in almost all the hypernuclei studied.
The small value of S$_\vlam$ and the substantial value for S$_\mathrm{N}$
mean that the effective LS and ALS interactions have to be of equal 
strength and opposite sign. The parametrization of the effective $\vlam$N 
interaction includes some three-body effects (see later) but, if
interpreted in terms of YN potential models, the value for $\vdel$ picks
out NSC97e,f~\cite{rijken99}. As noted in Sect.~\ref{sec:s-shell}
and Sect.~\ref{sec:L7Li}, these YN models have the correct
combination of spin-spin and $\vlam$--$\vsig$ coupling strengths
to account for data on \lamb{4}He (\lamb{4}{H}) and \lamb{7}{Li}.
They also have weak odd-state tensor interactions that give a
small positive value for $\mathrm{T}\sim 0.05$\,MeV. The LS
interaction, which gives rise to $\mathrm{S}_+\!=\!(\mathrm{S}_\vlam
+ \mathrm{S}_\mathrm{N})/2$, has roughly the correct strength but
the ALS interaction is only about one third as strong as the LS 
interaction, although with the correct relative sign. For the
newer ESC04 interactions~\cite{rijken06}, the ALS interaction is a little
stronger and the other components seem comparable to those of the favored
NSC97 interactions, except for differences in the odd-state central 
interaction. The attractive odd-state central interaction of the ESC04
models is favored by some data on $p_\vlam$ states over the 
overall repulsive interaction for the NSC97 models.
The most recent quark-model baryon-baryon interactions
of Fujiwara and collaborators~\cite{fujiwara04} also have trouble
explaining the small doublet splitting in \lamb{9}{Be}. 

 The mass dependence of the interaction parameters has been studied
by calculating the two-body matrix elements from YNG interactions
using Woods-Saxon wavefunctions. This approach requires the assignment
of binding energies for the p-shell nucleons and the $s_\varLambda$ orbit.
For the nucleons, binding energy effects are not so easy to deal with
because, as emphasized in Sect.~\ref{sec:pshell}, the nuclear
parentage is widely spread because of the underlying supermultiplet
structure of p-shell nuclei (the allowed removal of a nucleon
generally involves more than one symmetry [f] for the core and
states with different symmetries are widely separated in energy).
Perhaps the best that can be done is to take an average binding
energy derived from the spectroscopic centroid energy for the removed
nucleons. This changes rapidly for light systems up to $^8$Be, beyond
which it remains rather constant.

 Any description of the absolute binding energies of p-shell
hypernuclei (the B$_\vlam$ values) requires the consideration of
binding-energy effects and the introduction of three-body interactions,
real as in Fig.~\ref{fig:1} or effective from many-body theory
for a finite shell-model space. The empirical evidence for this
need is that the B$_\vlam$ values for p-shell hypernuclei don't grow 
as fast as $n\overline\mathrm{V}$, requiring a repulsive term quadratic 
in $n$ (the number of p-shell nucleons)~\cite{mgdd85}. Also, a 
description of $\vlam$ single-particle 
energies over the whole periodic table requires that the 
single-particle potential have a density dependence~\cite{millener88}, 
as might arise from the the zero-range three-body interaction in a 
Skyrme Hertree-Fock calculation. Much of the effect
of a three-body interaction is included in the parametrization of
the effective two-body $\vlam$N interaction. If two or one
s-shell nucleons are involved, the three-body interaction contributes
to the $\vlam$ single-particle energy or the effective 
two-body $\vlam$N interaction, respectively. This leaves only
the $p^2s_\vlam$ terms to consider. The real three-body
interactions derived from meson exchange present a problem for
shell-model calculations in that they possess singular short-range
behavior. In the two-body case, this is where the G-matrix or a purely
phenomenological treatment come in. For the three-body case, there
are too many independent matrix elements to parametrize, although
Gal, Soper, and Dalitz~\cite{gsd} have introduced a five parameter
representation of the two-pion-exchange three-body interaction.

 Given a set of three-body matrix elements, it is certainly possible
to include them in the shell-model calculations~\cite{gsd}. Another
useful extension of the shell-model codes would be to use the complete
$1\hbar\omega$ space for the non-normal-parity levels of p-shell
hypernuclei. Simple calculations for $p^np_\vlam$ configurations
have been done~\cite{auerbach83} but is important to include
the configurations involving  $1\hbar\omega$ states of the core
nucleus coupled to an $s_\vlam$. This is necessary to permit
the exclusion of spurious center-of-mass states from the shell-model
basis and to provide the amplitudes for nucleon emission leaving
low-lying states of the daughter hypernucleus with a $\vlam$ in
the $s_\vlam$ orbit, as indicated in Fig.~\ref{fig:lo16kpi}.
The configuration mixing also redistributes the strength from
from $p^np_\vlam$ states strongly formed in strangeness-exchange
or associated-production reactions. In calculating matrix elements
involving  the $p_\vlam$ orbit, it is important to include
binding-energy effects by using realistic radial wave functions because
the $p_\vlam$ orbit becomes bound only at $\mathrm{A}\sim 12$ and the
rms radius of the $\vlam$ orbit can be $\sim 4.5$\,fm compared with
$\sim 2.8$\,fm for the p-shell nucleon.

 The next generation of hypernuclear $\ugamma$-ray spectroscopy
experiments using a new Hyperball-J and the \Kpi reaction is being
prepared for J-PARC, starting perhaps early in 2009. The day-one
experiment~\cite{jparc} will be run at $p_\mathrm{K}\! =\! 1.5$\,GeV/c.
The spin-flip
amplitudes are strong in the elementary interaction between 1.1\,GeV/c
and 1.5\,GeV/c and the cross sections for spin-flip vs non-spin-flip
strength will be checked by using a $^4$He target and monitoring
the $\ugamma$ ray from the $1^+$ excited state of \lamb{4}{He}.
Also, the intention is to make a precise measurement of the lifetime
of the first-excited $3/2^+$ state of \lamb{7}{Li} using the
Doppler shift attenuation method. For \lam{10}{B}, the ground-state
doublet spacing will be determined unless it is smaller than 50\,keV.
For \lam{11}{B}, the power of a larger and more efficient detector
array will be used to sort out the complex level scheme by the use
of $\ugamma\ugamma$ coincidence measurements. Finally, a $^{19}$F
target will be used to measure the ground-state doublet spacing in
\lam{19}{F}.

 The measurement on \lam{19}{F} represents the start of a program
of $\ugamma$-ray spectroscopy on sd-shell nuclei. This will require
shell-model calculations for both $0\hbar\omega$ and $1\hbar\omega$
sd-shell hypernuclear states. In much of the first half of the
sd shell, supermultiplet symmetry, SU3 symmetry, and LS coupling
are still rather good symmetries. As a result, there are the same
opportunities as in the p shell to emphasize certain
spin-dependent components of the effective $\vlam$N interaction
by a judicious choice of target.
Now there are more two-body matrix elements -- 8 for $(sd)s_\varLambda$
-- and more sensitivity to the range structure of the $\vlam$N
effective interaction.

 The experiments just outlined represent the start of a very rich
experimental program using Hyperball-J at J-PARC (see
Ref.~\cite{hashtam06}). It should be possible to go to all the
way to rather heavy nuclei where the $p_\vlam$ orbit is below
the lowest particle-decay threshold (this is true for the special
case of \lam{13}{C}~\cite{kohri02}). 

\appendix

\section{Basics of Racah Algebra for SU2}
\label{sec:racsu2}

 The Wigner-Eckart theorem is used in the form~\cite{brink}
\begin{equation}
\langle J_fM_f|T^{kq}|J_iM_i\rangle = \langle J_iM_i\,kq|J_fM_f\rangle
\langle J_f||T^{k}||J_i\rangle \; .
\label{eq:wesu2}
\end{equation}

The elements of the unitary transformation that defines the recoupling
of three angular momenta are given by
\begin{equation}
 | \left[(j_1j_2)J_{12}j_3\right]J\rangle  = \sum_{J_{23}}
U(j_1j_2Jj_3,J_{12}J_{23})|\left[j_1(j_2j_3)J_{23}\right]J\rangle\; ,
\label{eq:racsu2}
\end{equation}
where the U-coefficient is simply related to the W-coefficient and
6j symbol~\cite{brink}
\begin{equation}
U(abcd,ef) = \widehat{e}\widehat{f}\,W(abcd,ef) =
\widehat{e}\widehat{f}(-)^{a+b+c+d}
\left\{ \begin{array}{ccc} a & b & e \\ d & c & f \end{array}\right\}\; .
\label{eq:ucofsu2}
\end{equation}

The elements of the unitary transformation that defines the recoupling
of four angular momenta are given by
\begin{eqnarray}
\lefteqn{ |\left[(j_1j_2)J_{12}(j_3j_4)J_{34}\right]J\rangle} \nonumber\\
 & & = \sum_{J_{13}J_{24}}
\pmatrix{
j_1 & j_2 & J_{12}  \cr
j_3 & j_4 & J_{34}  \cr
J_{13}  & J_{24} & J \cr}
 |\left[(j_1j_3)J_{13}(j_2j_4)J_{24}\right]J\rangle \; ,
\label{eq:9jsu2a}
\end{eqnarray}
where the normalized 9j symbol is simply related to the usual
9j symbol~\cite{brink}
\begin{equation}
\pmatrix{
a & b & c  \cr
d & e & f  \cr
g & h & i \cr}
= \widehat{c}\,\widehat{f}\,\widehat{g}\,\widehat{h}\,
\left\{ \begin{array}{ccc} a & b & c \\ d & e & f \\
 g & h & i\end{array}\right\} \; .
\label{eq:9jsu2b}
\end{equation}

The reduced matrix elements of a coupled operator consisting of
spherical tensor operators that operate in different spaces, e.g.
different shells or orbital and spin spaces, are given by
\begin{eqnarray}
\lefteqn{\langle J_1J_2;J||[R^{k_1},S^{k_2}]^k||J_1'J_2';J'\rangle }
\nonumber \\
 & & =
\pmatrix{
J_1' & k_1 & J_1  \cr
J_2' & k_2 & J_2  \cr
J'   & k & J \cr}
\langle J_1||R^{k_1}||J_1'\rangle\langle J_2||S^{k_2}||J_2'\rangle\; .
\label{eq:cprodsu2}
\end{eqnarray}

The reduced matrix elements of a coupled operator consisting of
spherical tensor operators that operate in the same space, e.g.
a coupled product of creation and annihilation operators acting
within the same shell, are given by
\begin{eqnarray}
\lefteqn{ \langle x\Gamma||[R^\sigma,S^\lambda]^\nu||x'\Gamma'\rangle }
\nonumber\\
 & & = (-)^{\sigma + \Lambda -\nu} \sum_{y\Gamma_1}
 U(\Gamma\sigma\Gamma'\lambda,\,\Gamma_1\nu)
\langle x\Gamma||R^\sigma ||y\Gamma_1\rangle
\langle y\Gamma_1|| S^\lambda||x'\Gamma'\rangle \; ,
\label{eq:intsu2}
\end{eqnarray}
where $\Gamma$ represents all the angular momentum type quantum numbers
such as JT or LST; $x$ and $y$ represent the other labels necessary
to specify the states spanning a space.

\section{Two-body Matrix Elements of the $\vlam$N Interaction}
\label{sec:twobod}

 Here, the two-body $p_\mathrm{N}s_\vlam$ matrix elements of the $\vlam$N
effective interaction in (\ref{eq:vlam}) are given in both LS and $jj$
coupling in terms of
the parameters $\overline\mathrm{V}$, $\vdel$, S$_\vlam$, S$_\mathrm{N}$,
and T from Table~\ref{tab:vlam}. Actually, the results are given in terms
of S$_+$ and S$_-$, the radial matrix elements associated with the
symmetric and antisymmetric spin-orbit interactions, respectively,
so that S$_\varLambda = \mathrm{S}_+ + \mathrm{S}_-$ and
S$_\mathrm{N} = \mathrm{S}_+ - \mathrm{S}_-$. Note that in
Ref.~\cite{gsd} the matrix elements are defined to give contributions
to the $\vlam$ binding energy B$_\vlam$ and that the order
of angular momentum coupling is (SL)J rather than (LS)J.
In LS coupling,
\begin{eqnarray}
\langle ^3P_0|V|^3P_0\rangle & = & \overline\mathrm{V} +\frac{1}{4}\varDelta
- 2\,\mathrm{S}_+ -6\,\mathrm{T} \nonumber \\
\langle ^3P_1|V|^3P_1\rangle & = & \overline\mathrm{V} +\frac{1}{4}\varDelta
- \mathrm{S}_+ +3\,\mathrm{T} \nonumber \\
\langle ^3P_2|V|^3P_2\rangle & = & \overline\mathrm{V} +\frac{1}{4}\varDelta
+ \mathrm{S}_+ -\frac{3}{5}\,\mathrm{T} \nonumber \\
 \langle ^1P_1|V|^1P_1\rangle & = & \overline\mathrm{V} -\frac{3}{4}\varDelta
\nonumber \\
 \langle ^3P_1|V|^1P_1\rangle & = &  - \sqrt{2}\,\mathrm{S}_- \; .
\label{eq:lstwo}
\end{eqnarray}
In $jj$ coupling,
\begin{eqnarray}
\langle p_{3/2}\ 1^-|V|p_{3/2}\ 1^-\rangle & = & \overline\mathrm{V} -\frac{5}{12}
\varDelta - \frac{1}{3}\mathrm{S}_+  +\mathrm{T} -\frac{4}{3}\mathrm{S}_-
\nonumber \\
\langle p_{3/2}\ 2^-|V|p_{3/2}\ 2^-\rangle & = & \overline\mathrm{V} +\frac{1}{4}
\varDelta + \mathrm{S}_+ - \frac{3}{5}\mathrm{T}  \nonumber \\
\langle p_{1/2}\ 0^-|V|p_{1/2}\ 0^-\rangle & = & \overline\mathrm{V} +\frac{1}{4}
\varDelta - 2\,\mathrm{S}_+ -6\,\mathrm{T}   \nonumber \\
\langle p_{1/2}\ 1^-|V|p_{1/2}\ 1^-\rangle & = & \overline\mathrm{V} -\frac{1}{12}
\varDelta - \frac{2}{3}\mathrm{S}_+ +2\,\mathrm{T}  + \frac{4}{3}\mathrm{S}_-
\nonumber \\
\langle p_{3/2}\ 1^-|V|p_{1/2}\ 1^-\rangle & = &  \frac{\sqrt{2}}{3}
\left\{ \varDelta - \mathrm{S}_+ -\mathrm{S}_- +3\,\mathrm{T}\right\} \; .
\label{eq:jjtwo}
\end{eqnarray}

\vspace{7mm}
This work has been supported by the US Department of Energy under Contract
No. DE-AC02-98CH10886 with Brookhaven National Laboratory.

\end{document}